%% file: iFTN.tex
\documentclass[aps,prx,twocolumn,superscriptaddress,a4paper,english,longbibliography]{revtex4-1}
\usepackage{times}
\usepackage{color}
\usepackage[colorlinks=true,urlcolor=blue,citecolor=blue]{hyperref}
\usepackage{url}
\usepackage{breakurl}
\usepackage{soul}
\usepackage{graphicx}
\usepackage{amsmath}
\usepackage{subfigure}
\usepackage{xcolor}
\usepackage{amssymb}
\usepackage{latexsym}
\usepackage{hyperref}
\DeclareGraphicsExtensions{.png,.eps}
\usepackage{float}
\usepackage{makecell}
\usepackage{diagbox}
\usepackage{setspace}
\usepackage{appendix}
\usepackage{etoolbox}

\newcommand {\R}{\textcolor {black}}

\usepackage{xcolor}
\usepackage[urlcolor=blue]{hyperref}      
\hypersetup{
	colorlinks = true,                    
	citecolor = {blue},
	linkcolor = {purple},
}

\begin{document}	
	
\title{Entanglement scaling and criticality of infinite-size quantum many-body systems in continuous space addressed by a tensor network approach} 

\author{Rui Hong}
\affiliation{Center for Quantum Physics and Intelligent Sciences, Department of Physics, Capital Normal University, Beijing 10048, China}
\author{Hao-Wei Cui}
\affiliation{Center for Quantum Physics and Intelligent Sciences, Department of Physics, Capital Normal University, Beijing 10048, China}
\author{An-Chun Ji}
\affiliation{Center for Quantum Physics and Intelligent Sciences, Department of Physics, Capital Normal University, Beijing 10048, China}
\author{Shi-Ju Ran}\email[Corresponding author. Email: ]{sjran@cnu.edu.cn}
\affiliation{Center for Quantum Physics and Intelligent Sciences, Department of Physics, Capital Normal University, Beijing 10048, China}

\date{\today}
\begin{abstract}
Simulating strongly-correlated quantum systems in continuous space belongs to the most challenging and long-concerned issues in quantum physics. This work investigates the quantum entanglement and criticality of the ground-state wave-functions of infinitely-many coupled quantum oscillators (iCQOs). The essential task involves solving a set of partial differential equations (Schrödinger equations in the canonical quantization picture) with infinitely-many variables, which currently lacks valid methods. By extending the imaginary-time evolution algorithm with translationally-invariant functional tensor network, we simulate the ground state of iCQOs with the presence of two- and three-body couplings. We determine the range of coupling strengths where there exists a real ground-state energy (dubbed as physical region). With two-body couplings, we reveal the logarithmic scaling law of entanglement entropy (EE) and the polynomial scaling law of correlation length against the virtual bond dimension $\chi$ at the dividing point of physical and non-physical regions. These two scaling behaviors are signatures of criticality, according to the previous results in quantum lattice models, but were not reported in continuous-space quantum systems. The scaling coefficients result in a central charge $c=1$, indicating the presence of free boson conformal field theory (CFT). We further show that the presence of three-body couplings, for which there are no analytical or numerical results, breaks down the CFT description at the dividing point. Our work uncovers the scaling behaviors of EE in the continuous-space quantum many-body systems. These results provide solid numerical evidence supporting the high efficiency of TN in representing the continuous-space quantum many-body wave-functions in the thermodynamic limit, and meanwhile suggest an efficient approach to study the entanglement properties and criticality in continuous space. 
\end{abstract}
\maketitle

\section{Introduction}
Developing efficient approaches for simulating strongly-correlated quantum systems is a long-standing topic of interest. A well-established treatment is lattice approximation that leads to quantum lattice models such as Heisenberg and Hubbard models, where tensor network (TN) methods~\cite{TNs3,TNs1,TNs2,TNbook} have been successfully adopted to reveal fruitful exotic phenomena such as quantum criticality and quantum topological orders~\cite{toporder0,toporder1,toporder2,toporder3,TNCriticality1,TNCriticality2}. The success of TN methods in simulating quantum lattice models is solidly supported by the scaling theories of entanglement entropy (EE)~\cite{entropy,mps,Scaling,arealaw2,arealaw1,critical1,mps_scaling}. For instance, matrix product state (MPS)~\cite{mpschushi1,mpschushi2} can faithfully represent the states satisfying one-dimensional (1D) area law of EE, which include the ground states of gapped 1D Hamiltonians with local couplings~\cite{mps,TNmps,gapped1d}. For 1D critical quantum lattice models, the criticality (such as central charge) can be accurately accessed by the scaling behaviors of MPS~\cite{mps_scaling, entropy, Scaling, mps}. 

Without lattice approximation, the main challenge lies in solving the many-body Schrödinger equations, which are multi-variable partial differential equations (PDE's). Analytical solutions are nearly infeasible. Numerical methods, such as Monte Carlo~\cite{mc2,qmk0,mk2,qmk} and neural-network schemes~\cite{NN4,NN3,NN1,NNa,NNb}, are severely hindered by the so-called ``exponential wall'', namely the complexity (say the dimension of sampling space) scales exponentially with the system size. Meanwhile, theories and methods established for quantum lattice models, such as EE and its scaling laws, can hardly be applied to the quantum models in continuous space. These severely hinder the investigation of the realistic quantum systems such as the quantum materials with strong correlations.


\begin{figure}[tbp]
	\centering
	\includegraphics[angle=0,width=0.9\linewidth]{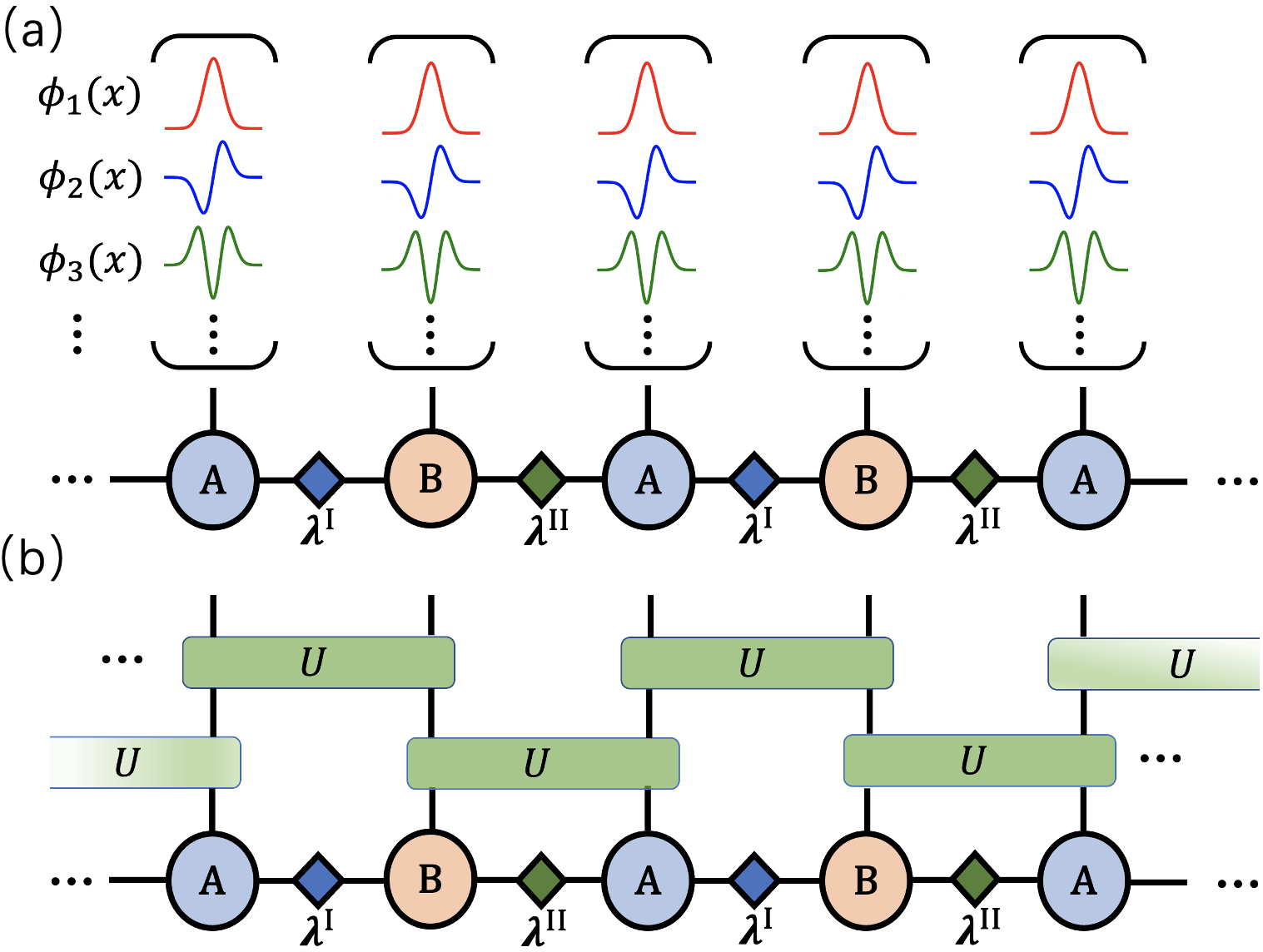}
	\caption{(Color online) (a) The illustration of iTEBD algorithm incorporated with the two-tensor translationally-invariant function MPS that represents the coefficients of a quantum wave-function expanded in a set of orthonormal functional bases $\{\phi_{s}(x)\}$. (b) The illustration of the imaginary-time evolution in the iTEBD algorithm, where $\hat{U}$ denotes the local evolution gate. Note the MPS and its time evolution can be readily generalized to the $K$-tensor translationally-invariant cases for $K >2$.}
	\label{fig-flow}
\end{figure}

In this work, we extend the TN-based imaginary-time evolution algorithms for quantum lattice models~\cite{itebd,itebd1} to the simulation of continuous-space quantum many-body systems by incorporating the continuous-space TN method~\cite{mpscon, FTN} (see Fig.~\ref{fig-flow}). We here adopt the translationally-invariant MPS to solve the time-independent Schrödinger equation (in the canonical quantization form) of infinitely-many coupled quantum oscillators (iCQOs). This Schrödinger equation consists of a set of PDE's with infinitely-many variables. No numerical results of solving these PDE's were reported to our best knowledge, due to the infinite number of variables involved. \R{One may refer to Refs.~[\onlinecite{MathMPS1,MathMPS2,MathMPS3}] on solving PDE's with finite variables by MPS.} The translationally-invariant MPS is used to represent the infinitely-many coefficients of the wave-function in a set of orthonormal functional bases.

By leveraging the advantages of canonical MPS~\cite{itebd} on efficiently simulating entanglement and correlations of quantum many-body states, we investigate the criticality of iCQOs with the presence of two- and three-body coupling terms. With just the two-body terms, we determine the range of coupling strength where there exists a physical ground-state solution (i.e., real ground-state energy as the Hamiltonian is hermitian~\cite{coupled}), which we dub as physical region. At the dividing point between the physical and non-physical regions, we uncover that the ground-state EE and correlation length satisfy the critical scaling laws (against the virtual dimension of MPS), which were previously uncovered in the critical quantum lattice models~\cite{entropy,corr,critical1,Scaling}. The scaling coefficients result in a central charge $c=1$~\cite{field_theory,harmonicCFT}, indicating that the ground state of iCQOs can be described by free boson conformal field theory (CFT)~\cite{CFT1}. With three-body couplings, we show the breakdown of CFT even with very weak three-body couplings. Our work explicitly shows the satisfaction of the 1D scaling laws of EE, which can serve as solid supports on MPS as an efficient ansatz to represent the ground-state wave-functions of continuous-space many-body systems, including those in the thermodynamic limit.

\section{Imaginary-time evolution for the ground-state simulation of infinitely-many coupled quantum oscillators}
To simulate the continuous-space quantum many-body systems, we expand the many-body wave-function, which is a normalized many-variable function denoted as $\psi(\boldsymbol{x})$ with $\boldsymbol{x} = (x_{1}, x_{2}, \ldots)$, in a set of complete functional bases $\{\phi_{s_{n}} (x_{n})\}$. A key idea is to encode the coefficients, which are exponentially many, in an MPS whose complexity scales just linearly with the system size (i.e., number of variables)~\cite{FTN, mpscon}. The infinite functional MPS is written as
\begin{eqnarray}
\psi(\boldsymbol{x}) = \operatorname{tTr} && \bigg( \prod_{n=1}^{\infty} A_{s_{2n-1} a_{2n-1} a_{2n}} B_{s_{2n} a_{2n} a_{2n+1}} \nonumber \\ && \lambda_{a_{2n}}^{\text{I}} \lambda_{a_{2n+1}}^{\text{II}} \phi_{s_{2n-1}} (x_{2n-1}) \phi_{s_{2n}} (x_{2n}) \bigg),
\label{eq-fmps}
\end{eqnarray}
with $\operatorname{tTr}$ the summation of all shared indexes. We take the Planck constant $\hbar=1$ for simplicity. We here choose the functional bases to be ``local'', i.e., constructed by the single-variable functions $\{\phi_{s_{n}} (x_{n})\}$ that satisfy $\int_{-\infty}^{\infty} \phi^{\ast}_{s} (x) \phi_{s'} (x) dx = \delta_{ss'}$. The physical dimension (the bond dimension of physical index $s_{n}$, denoted as $D$) determines the order of expansion, namely $\dim(s_{n})$ in $\{\phi_{s_{n}} (x_{n})\}$. The virtual dimension (the dimension of virtual index $a_{n}$, denoted as $\chi$) determines the number of expansion terms. Both dimensions determine the parameter complexity of the MPS.

To efficiently represent the infinite-size wave-functions, we take the MPS to be translationally invariant. Fig.~\ref{fig-flow} illustrates the MPS satisfying the ``two-tensor'' translational invariance, namely the MPS is formed by infinitely-many copies of two inequivalent tensors $\boldsymbol{A}$ and $\boldsymbol{B}$, as well as those of two inequivalent spectra (diagonal matrices) $\lambda^{\text{I}}$ and $\lambda^{\text{II}}$ [Fig.~\ref{fig-flow}(a)]. The spectra are necessary to define the canonical form of infinite MPS for the purpose of optimal truncations and entanglement simulations~\cite{itebd}. The number of variational parameters in such an MPS scales approximately as $O(D\chi^{2})$. In practice, we adopt the ``$K$-tensor'' translational invariance. We take $K=2$ when there exist only the one- and two-body terms in the Hamiltonian, and $K=6$ with the presence of the three-body terms.

Translational invariance allows TN methods to efficiently simulate infinite-size quantum many-body systems. For quantum lattice models, translational invariance explicitly corresponds to the translational invariance of the quantum states with the respect to lattice sites, since the tensors are usually in one-to-one correspondence with the sites. For continuous-space systems (such as iCQOs considered here), there is no such correspondences. The translational invariance respects the symmetry of the Hamiltonian, and is not contradictory to the indistinguishability of particles. Thus, we regard translational invariance as a physically-reasonable conjecture imposed on the functional MPS ansatz. Note that the Schrödinger equations we aim to solve are in the canonical quantization picture, where the bosonic or fermionic statistics of the wave-functions is not explicitly considered. 

The Hamiltonian of iCQOs is written as 
\begin{eqnarray}
\hat{H} = &&-\frac{1}{2}\sum_{n=1}^{\infty} (\frac{\partial^{2}}{\partial x_{n}^{2}} - \omega^{2} x_{n}^{2}) + \gamma \sum_{n=1}^{\infty} x_{n} x_{n+1},
\label{eq-hamilt}
\end{eqnarray}
where we take the natural frequency $\omega=1$ for simplicity. The time-independent Schrödinger equation reads $\hat{H}|\psi\rangle = E |\psi\rangle$, which is a set of PDE's consisting of infinitely-many variables $\{x_{n}\}$. The minimal eigenvalue (denoted as $E_{0}$) gives the ground-state energy, satisfying $E_{0} = \min_{|\psi\rangle} \langle \psi | \hat{H} | \psi \rangle = \langle \psi_{0} | \hat{H} | \psi_{0} \rangle $ with $|\psi_{0}\rangle$ denoting the ground state.

Given a set of orthonormal functional bases $\{\phi_{s}(x_{n})\}$, the operators (say $\frac{\partial}{\partial x_{n}}$ and $x_{n}$) can be represented as matrices (denoted as $\hat{\partial}_{n}$ and $\hat{X}_{n}$). Their coefficients satisfy $\partial_{ss'} = \int_{-\infty}^{\infty} \phi^{\ast}_{s}(x) \frac{\partial}{\partial x} \phi_{s'}(x) dx$ and $X_{ss'} = \int_{-\infty}^{\infty} \phi^{\ast}_{s}(x) x  \phi_{s'}(x) dx$. We then re-write the Hamiltonian as $\hat{H} = \sum_{n=1}^{\infty} \hat{H}_{n,n+1}$ with $\hat{H}_{n,n+1} = -\frac{1}{4}(\hat{\partial}^{2}_{n} + \hat{\partial}^{2}_{n+1} - \hat{X}^{2}_{n} - \hat{X}^{2}_{n+1}) + \gamma \hat{X}_{n} \hat{X}_{n+1}$.

We adopt the infinite time-evolving block decimation (iTEBD) algorithm~\cite{itebd,itebd1} to simulate the ground state of iCQOs. We may define the two-body imaginary-time evolution gate \R{$\hat{U}_{n} = e^{-\tau \hat{H}_{n,n+1}}$} with $\tau$ a small positive number called Trotter step. By iteratively implementing the gates $\{\hat{U}_{n}\}$ on an initial state $|\psi_{\text{init}}\rangle$ [see Fig.~\ref{fig-flow}(b)], one has $\frac{1}{Z}\lim_{K \to \infty} (\prod_{n}\hat{U}_{n})^{K} = |\psi_{0}\rangle$ with $Z$ a normalization factor. The evolution scheme can be readily generalized to the cases with $K$-site translational invariance, in order to handle the simulations with multi-body couplings. More details can be found in the Supplemental Material~\cite{SM}.

\section{Numerical results of critical scaling laws}
\begin{figure}[tbp]
	\centering
	\includegraphics[angle=0,width=1\linewidth]{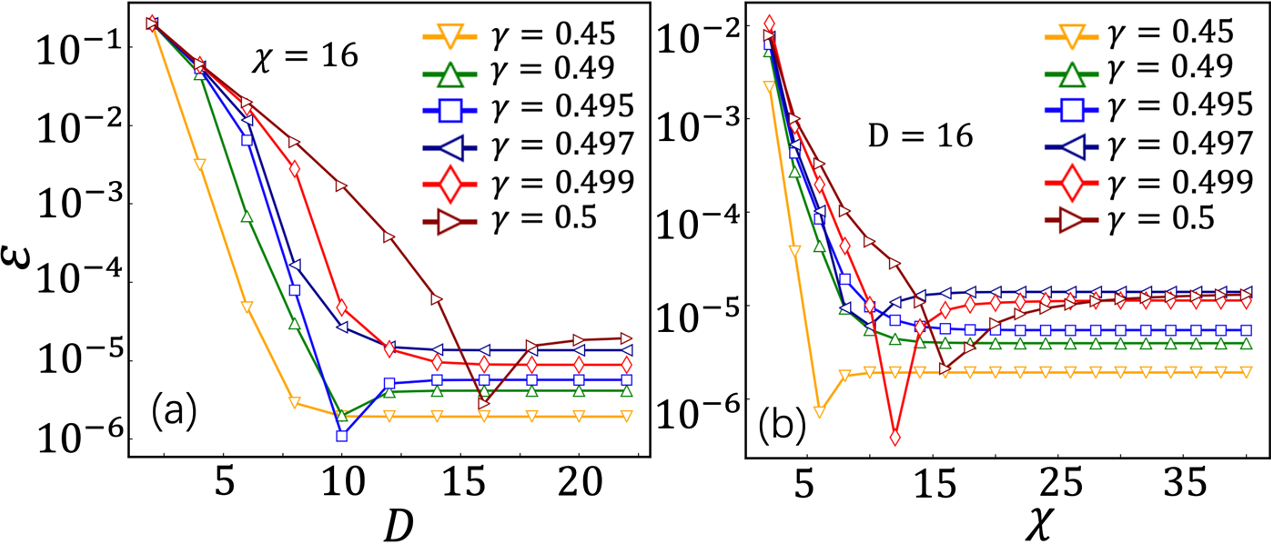}
	\caption{(Color online) (a) The error of the ground-state energy $\varepsilon$ versus the expansion order $\mathcal{D}$ (physical dimension) while fixing the virtual bond dimension $\chi = 16$. (b) the error of the ground-state energy $\varepsilon$ versus the virtual index $\chi$ by fixing $\mathcal{D}=16$. We vary the strength of the two-body interaction from ${\gamma} = 0.45$ to $0.5$.}
	\label{fig-2Derror}
\end{figure}

The ground-state energy of iCQOs with two-body couplings [Eq.~(\ref{eq-hamilt})] can be analytically obtained~\cite{coupled}, which satisfies
\begin{eqnarray}
	\R{E_{\text{exact}}=\lim_{N \to \infty} \frac{1}{2}\sum_{n=1}^{N}\sqrt{1+2\gamma \cos\left(\frac{n\pi}{N+1} \right)}.}
	\label{eq-exactE}
\end{eqnarray}
$E_{\text{exact}}$ is real for $\gamma< = \gamma_{c} \equiv 0.5$, which we dub as the physical region. Fig.~\ref{fig-2Derror} shows the error $\varepsilon = |E_{\text{exact}} - \langle \psi|\hat{H} |\psi \rangle|$ versus the physical and virtual dimensions ($D$ and $\chi$). The error converges to $\varepsilon \sim O(10^{-5})$, indicating that the ground states can be well approximated by the MPS's with a finite $D$ or $\chi$.

\begin{figure}[tbp]
	\centering
	\includegraphics[angle=0,width=1\linewidth]{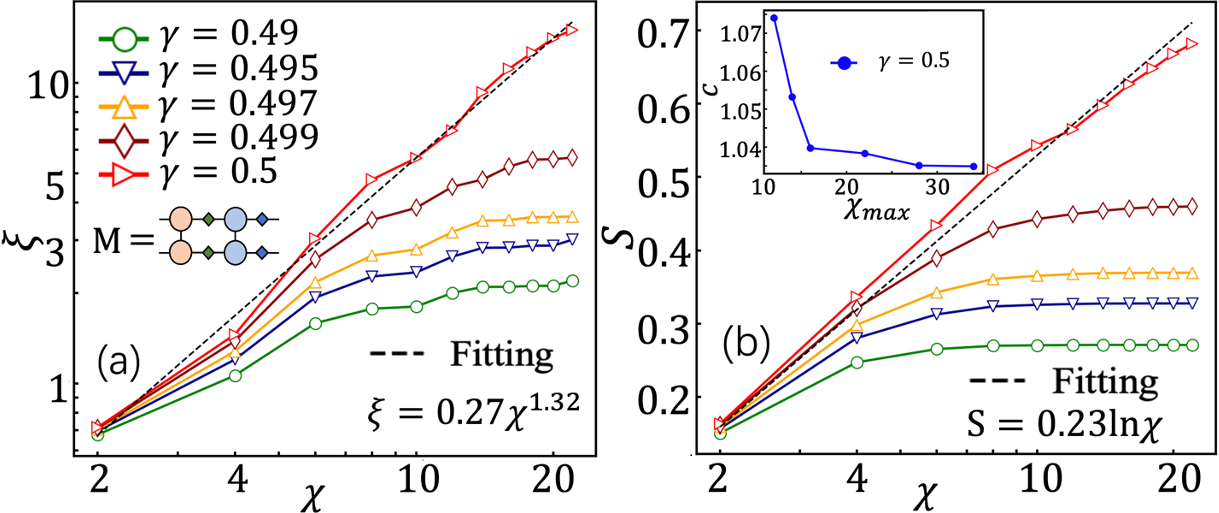}
	\caption{(Color online) (a) The correlation length $\xi$ and (b) EE $S$ versus the virtual bond dimension $\chi$. We vary the strength of the two-body interaction from ${\gamma} = 0.49$ to $0.5$ by fixing $\mathcal{D}=16$. The correlation length is computed using the transfer matrix of the infinite MPS illustrated in the inset of (a). The inset of (b) shows the convergence of central charge $c \to 1$ against $\chi_{max}$ (meaning the fitting is implemented with the data for $\chi \leq \chi_{max}$). Fittings of $\xi$ and $S$ are given by the dashed lines, which obey algebraic and logarithmic scaling laws, respectively [see Eqs. (\ref{eq-xi}) and (\ref{eq-ent})].}
	\label{fig-2Dentropycorr}
\end{figure}

The correlation length $\xi$ and entanglement entropy (EE) $S$ of the ground state can be efficiently accessed by the MPS method~\cite{TNbook, mpsreview}. The correlation length is obtained from the eigenvalues of the transfer matrix $\boldsymbol{M}$ of the infinite MPS [see the inset of Fig.~\ref{fig-2Dentropycorr}(a) for the illustration of $\boldsymbol{M}$], satisfying $\xi = \frac{1}{\ln \Lambda_{0} - \ln \Lambda_{1}}$ with $\Lambda_{0}$ and $\Lambda_{1}$ the two dominant eigenvalues of  $\boldsymbol{M}$. It naturally obeys the standard definition of correlation length of continuous-space wave-functions, which characterizes the decay of correlation function $\langle \hat{O}_{n} \hat{O}_{n'} \rangle$ versus $|n-n'|$. The EE of continuous-space wave-functions is much less investigated. Here, the canonical form of the infinite MPS gives us a well-defined entanglement spectrum. The EE can then be defined as $S = -2\sum_a (\lambda_a^{\text{I} (\text{II})})^2\ln \lambda_a^{\text{I}(\text{II})}$. The difference between the EE's calculated from the two spectra is about $O(10^{-3})$, which is the result of numerical errors.

When $\gamma$ is far away from the dividing point $\gamma=0.5$, both the correlation length $\xi$ and EE $S$ converge as $\chi$ increases (Fig.~\ref{fig-2Dentropycorr}). We fix $\mathcal{D}=16$, which should be sufficiently large. These results indicate the finiteness of correlation length and EE for the ground state of iCQOs within the physical region, which are typical signatures of a finite excitation gap. This suggests that an infinite functional MPS with finite bond dimensions can well approximate such non-critical many-body wave-functions.

As $\gamma$ approaches the dividing point, both $\xi$ and $S$ become asymptotically divergent (Fig.~\ref{fig-2Dentropycorr}). At $\gamma=0.5$, our results suggest the algebraic scaling law of correlation length as
\begin{eqnarray}
	\xi \sim \chi^{\kappa},
	\label{eq-xi}
\end{eqnarray}
and meanwhile the logarithmic scaling law of EE as
\begin{eqnarray}
	S \sim \eta \ln \chi.
	\label{eq-ent}
\end{eqnarray}
We have $\kappa \simeq 1.32 $ and $\eta \simeq 0.23$. These two scaling laws are recognized as signatures of criticality for quantum lattice models~\cite{Scaling}, but haven't been uncovered in the continuous-space systems such as iCQOs. Our work reveals that even with divergent correlation length and EE, the continuous-space wave-functions can still be asymptotically approached by MPS in a similar manner for approaching the critical quantum lattice systems.

The critical properties can be further characterized by the scaling coefficients. From CFT, we can estimate the central charge as
\begin{eqnarray}
	c = \frac{6\eta}{\kappa} = 1 + O(10^{-2}).
	\label{eq-central}
\end{eqnarray}
This suggests the wave-function at the dividing point between the physical and non-physical regions to be described by the $c=1$ boson CFT~\cite{CFT1}. The inset of Fig.~\ref{fig-2Dentropycorr}(b) demonstrates the convergence of $c$ versus $\chi$.

\section{Breakdown of criticality with three-body couplings}
\begin{figure}[tbp]
	\centering
	\includegraphics[angle=0,width=1\linewidth]{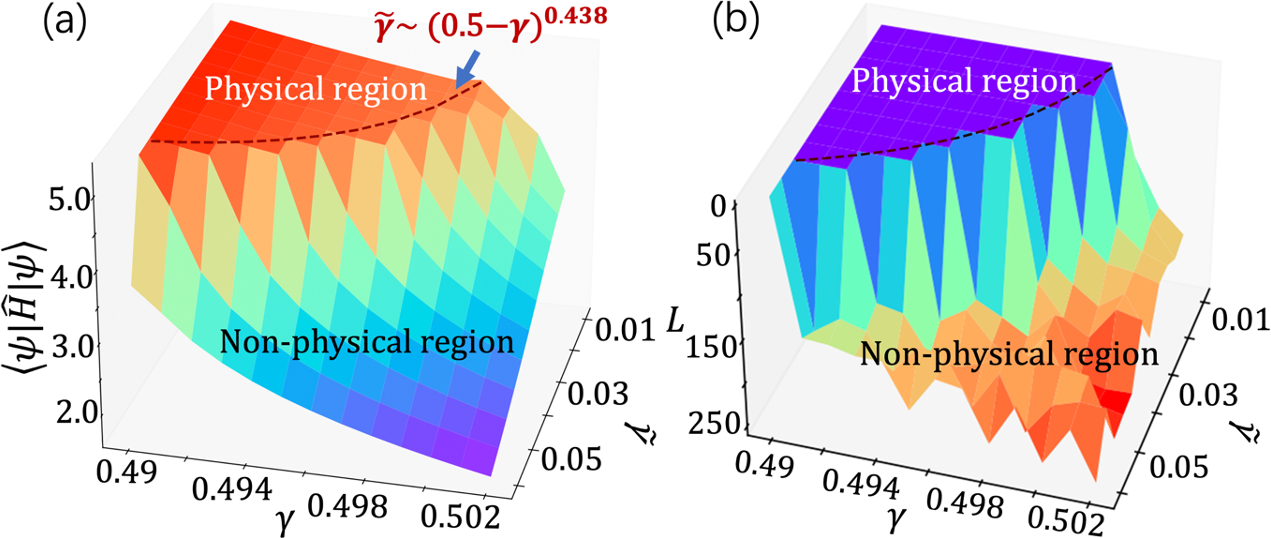}
	\caption{(Color online) (a) The variational ground-state energy $\langle \psi |\hat{H}| \psi \rangle$ and (b) the violation of Schrödinger equation $L$ [Eq.~(\ref{eq-eqL})] with different two- and three-body coupling strengths ($\gamma$ and $\tilde{\gamma}$, respectively). The dividing boundary between the physical and non-physical regions obeys $\tilde{\gamma} =0.896 \left( 0.5 - \gamma \right)^{0.438}$.}
	\label{fig-3Dphase}
\end{figure}

Below, we introduce the three-body coupling terms $\tilde{\gamma} \sum_{n} x_{n}x_{n+1}x_{n+2}$ to the Hamiltonian with $\tilde{\gamma}$ the coupling strength. Fig.~\ref{fig-3Dphase} show the physical region determined by the variational ground-state energy $\langle \psi|\hat{H}|\psi\rangle$ and the violation of Schrödinger equation
\begin{eqnarray}
	L = \left|\hat{H} |\psi \rangle-E|\psi \rangle \right|^2,
	\label{eq-eqL}
\end{eqnarray}
with $E = \langle \psi |\hat{H}|\psi\rangle$. In the non-physical region, the obtained energy is irregularly low with the violation $L \sim O(10^{2})$ (which is far larger than the violation $L \sim O(10^{-1})$ in the physical region). By fitting, we find the dividing boundary to satisfy an algebraic relation $\tilde{\gamma} \sim \left( 0.5 - \gamma \right)^{0.438}$.

\begin{figure}[tbp]
	\centering
	\includegraphics[angle=0,width=1\linewidth]{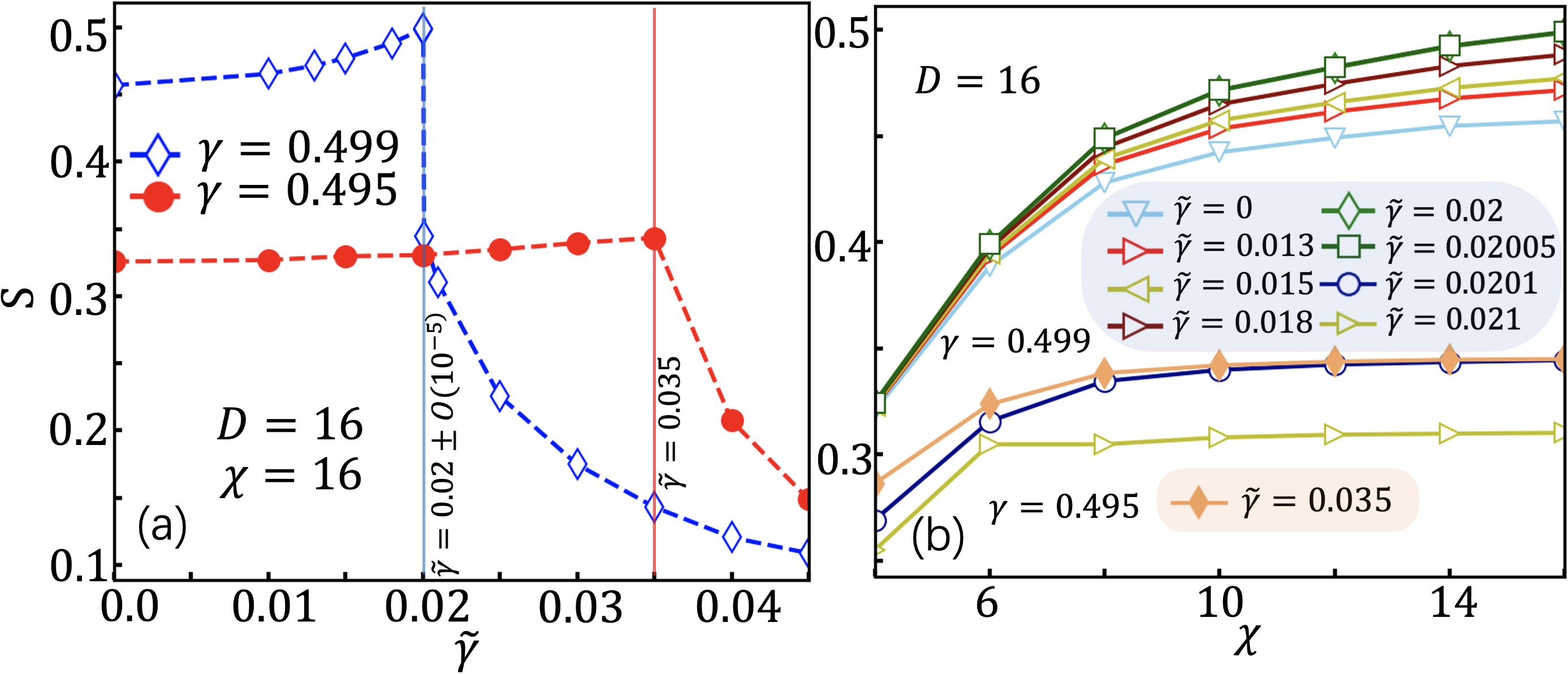}
	\caption{(Color online) (a) The ground-state EE $S$ versus the strength of three-body couplings $\tilde{\gamma}$ for $\gamma=0.499$ and $0.495$. We fix $D=16$ and $\chi=16$. (b) The scaling of EE against $\chi$. We take different strengths of three-body couplings $\tilde{\gamma}$ for $\gamma=0.499$, and take $\tilde{\gamma} = 0.035$ for $\gamma=0.495$.}
	\label{fig-3Dentropy}
\end{figure}

Fig.~\ref{fig-3Dentropy}(a) shows that EE slightly increases when $\tilde{\gamma}$ approaches the dividing boundary from the physical side. The dividing boundary given by the peak of EE is consistent with that obtained in Fig.~\ref{fig-3Dphase}. We have $\tilde{\gamma} = 0.02 \pm O(10^{-5})$ as the dividing point for $\gamma = 0.499$, and $\tilde{\gamma} = 0.035 \pm O(10^{-5})$ for $\gamma = 0.495$. On the dividing boundary, our results show that EE no longer obeys the logarithmic scaling law but converges after $\chi \simeq 25$ for $\gamma = 0.499$ (or after $\chi \simeq 15$ for $\gamma = 0.495$). See Fig.~\ref{fig-3Dentropy}(b). EE does not scale to diverge as $\chi$ increases, and thus implies the absence of criticality or the CFT description for the ground-state wave-functions when the three-body terms appear.

\section{Summary}
This work concerns simulating the ground-state wave-functions and their entanglement properties of infinite-size quantum many-body systems in continuous space. We combine the infinite time-evolving block decimation algorithm with functional tensor network (TN) to simulate the ground states of infinite coupled quantum oscillators (iCQOs) defined in continuous space. At the dividing point between the physical and non-physical regions with just two-body couplings, we reveal algebraic scaling laws of the correlation length and logarithmic scaling laws of the entanglement entropy (EE) against the virtual dimension of the TN ansatz. These findings suggest the $c=1$ boson conformal field theory (CFT) describing the criticality at the dividing point. We further show that the three-body terms break down the criticality and the CFT description on the dividing boundary. Our work demonstrates the validity of EE scaling theories based on TN for investigating the quantum many-body systems in continuous space. The satisfactions of EE scaling laws serves as a solid evidence on the efficiency of TN for approximating continuous-space wave-functions and for accessing the properties of criticality of continuous-space quantum many-body systems.

\section*{Acknowledgment}  R.H. is grateful to Ding-Zu Wang for helpful discussions.This work was supported in part by the Beijing Natural Science Foundation (Grant No. 1232025), NSFC (Grant
No. 62175169), the Ministry of Education Key Laboratory of Quantum Physics and Photonic Quantum Information (Grant No. ZYGX2024K020), and the Academy for Multidisciplinary Studies, Capital Normal University. S.J.R. acknowledges the support from the Peng Huanwu Visiting Professor Program, the Chinese Academy of Sciences. The machine learning simulations were partially performed on the robotic AI-Scientist platform of Chinese Academy of Sciences.

\section*{DATA AVAILABILITY} The data that support the findings of this work are available from the corresponding author upon reasonable request.

\appendix
\section{Functional matrix product state and operators}

To simulate the continuous-space quantum many-body systems, we expand the many-body wave-function, which is a normalized many-variable function denoted as $\psi(\boldsymbol{x})$ with $\boldsymbol{x} = (x_{1}, x_{2}, \ldots)$, in a set of orthonormal functional bases $\{\phi_{s_{n}} (x_{n})\}$. A key idea is to encode the coefficients, which are exponentially many, in a tensor network (TN) whose complexity scales just polynomially with the system size~\cite{mpscon,FTN}. 

Here, we take the matrix product state (MPS), where the many-body wave-function can be expressed as
 \begin{equation}
	\psi(\mathbf{x}) = \sum_{s_1\cdots s_N=0}^{\mathcal{D}-1} C_{s_1\cdots s_N} \phi_{s_1}(x_1)\cdots\phi_{s_N}(x_N).
	\label{eq-mfun}
\end{equation}

Below, we adopt the infinite MPS, which is a simple but powerful TN to represent infinite-size quantum states. To represent many-body wave-functions in the continuous space, the infinite functional MPS can be written as
\begin{eqnarray}
\psi(\boldsymbol{x}) = \operatorname{tTr} && \bigg( \prod_{n=1}^{\infty} A_{a_{2n-1} s_{2n-1} a_{2n}} \lambda_{a_{2n}}^{\text{I}} B_{a_{2n} s_{2n} a_{2n+1}} \nonumber \\ && \lambda_{a_{2n+1}}^{\text{II}} \phi_{s_{2n-1}} (x_{2n-1}) \phi_{s_{2n}} (x_{2n}) \bigg),
\label{eq-fmps}
\end{eqnarray}

with $\operatorname{tTr}$ denoting the summation over all shared indexes. The indexes $\alpha_{n}$ ($n=1, \cdots, N-1$) are dubbed as virtual bonds. The upper bound of their dimensions $\dim(\alpha_{n})$ is called the virtual bond dimension, denoted by $\chi$. The indexes $\{s_{n}\}$ are called the physical bonds, with their dimensions called the physical bond dimension , denoted by $\mathcal{D}$. In our cases for representing continuous-space wave-functions, $\mathcal{D}$ determines the expansion order in the functional bases. The number of parameters in the MPS (i.e., the total number of elements in the tensors $\{\mathbf{A}^{(n)}\}$ for $n=1, \cdots, N$) scales only linearly with $N$ as $O(N\mathcal{D}\chi^{2})$, while the number of coefficients in $\psi(\mathbf{x})$ scales exponentially as $O(\mathcal{D}^{N})$. 

The MPS is formed by infinite copies of four inequivalent tensors, which are third-order tensors $\boldsymbol{A}$, $\boldsymbol{B}$, and vectors (or diagonal matrices) $\boldsymbol{\lambda}^{\text{I}}$, $\boldsymbol{\lambda}^{\text{II}}$. Such an MPS is called two-site or two-tensor (considering $\boldsymbol{A}$ and $\boldsymbol{B}$ are defined on the sites) translationally invariant. 

To solve the Schr\"odinger equation for the ground state, we need to obtain the coefficients of operators in the same functional bases. Considering an operator $\hat{O}^{(m)}$ with respect to the variable $x_{m}$, acting it on $\psi(\mathbf{x})$ yields
\begin{eqnarray}
\tilde{\psi}(\boldsymbol{x}) &=& \hat{O}^{(m)} [\psi(\boldsymbol{x})] \nonumber \\ 
&=& \sum_{s_1\cdots s_N=0}^{\mathcal{D}-1} C_{s_1\cdots s_N} [\prod_{n \neq m} \phi_{s_n}(x_n)] \nonumber \\ && \hat{O}^{(m)}[\phi_{s_m}(x_m)].
\label{eq-Opsi}
\end{eqnarray}
Since the functional bases are assumed to be orthonormal, we have the expansion
\begin{eqnarray}
\hat{O}^{(m)}[\phi_{s}(x)] = \sum_{s'=0}^{\mathcal{D}-1} O^{(m)}_{ss'} \phi_{s'}(x),
\label{eq-Oss}
\end{eqnarray}
where $O^{(m)}_{ss'}$ denotes the coefficients of $\hat{O}^{(m)}$. Substituting it into Eq.~(\ref{eq-Opsi}), we have
\begin{eqnarray}
\tilde{\psi}(\boldsymbol{x}) &=& \sum_{s_1\cdots s_N=0}^{\mathcal{D}-1} C_{s_1\cdots s_N} [\prod_{n \neq m} \phi_{s_n}(x_n)] \sum_{s'_{m}=0}^{\mathcal{D}-1} O^{(m)}_{s_{m}s'_{m}} \phi_{s'_{m}}(x_{m}) \nonumber \\ 
&=& \sum_{s_1\cdots s_N=0}^{\mathcal{D}-1} \sum_{s'_{m}=0}^{\mathcal{D}-1} O^{(m)}_{s_{m}s'_{m}} C_{s_1\cdots s_N}\nonumber \\ &&  [\prod_{n \neq m} \phi_{s_n}(x_n)] \phi_{s'_{m}}(x_{m}).
\label{eq-Opsi1}
\end{eqnarray}

Rearranging the above equation (say, with the replacement $s_{m} \leftrightarrow s'_{m}$), we have
\begin{eqnarray}
\tilde{\psi}(\boldsymbol{x}) =  \sum_{s_1\cdots s_N=0}^{\mathcal{D}-1}  (\sum_{s'_{m}=0}^{\mathcal{D}-1}  O^{(m)}_{s'_{m}s_{m}} \nonumber \\ && C_{s_1\cdots s_{m-1} s'_{m} s_{m+1} s_N})  \prod_{n} \phi_{s_n}(x_n).
\label{eq-Opsi2}
\end{eqnarray}
Denote the expansion of $\tilde{\psi}(\boldsymbol{x})$ as $\tilde{\psi}(\mathbf{x}) = \sum_{s_1\cdots s_N=0}^{\mathcal{D}-1} \tilde{C}_{s_1\cdots s_N} \phi_{s_1}(x_1)\cdots\phi_{s_N}(x_N)$ with $\tilde{\boldsymbol{C}}$ the coefficient tensor, Then, comparing with Eq.~(\ref{eq-Opsi2}), we immediately have
\begin{eqnarray}
\tilde{C}_{s_1\cdots s_N} = \sum_{s'_{m}} O^{(m)}_{s'_{m}s_{m}} C_{s_1\cdots s_{m-1} s'_{m} s_{m+1} s_N}.
\label{eq-Ctilde}
\end{eqnarray}

When the coefficient tensor $\boldsymbol{C}$ is represented as an MPS, the contraction between $\boldsymbol{O}^{(m)}$ and $\boldsymbol{C}$ becomes the contraction between $\boldsymbol{O}^{(m)}$ and the corresponding local tensor $\boldsymbol{A}^{(m)}$ as
\begin{eqnarray}
\tilde{A}^{(m)}_{\alpha s \alpha'} = \sum_{s'} O^{(m)}_{s's} A^{(m)}_{\alpha s' \alpha'}.
\label{eq-MPStensorO}
\end{eqnarray}
Replacing $\boldsymbol{A}^{(m)}$ by  $\boldsymbol{\tilde{A}}^{(m)}$, one obtains all local tensors of the MPS representing $\tilde{\psi}(\boldsymbol{x})$.

To evaluate $\boldsymbol{O}^{(m)}$, let us return to Eq.~(\ref{eq-Oss}) and perform the following integral
\begin{eqnarray}
\int_{-\infty}^{\infty} \phi^*_{s'}(x) \hat{O}^{(m)}[\phi_{s}(x)] dx 
&=& \int_{-\infty}^{\infty} \phi^*_{s'}(x) \sum_{s''=0}^{\mathcal{D}-1} O^{(m)}_{ss''} \phi_{s''}(x) dx  \nonumber \\
&=& \sum_{s''=0}^{\mathcal{D}-1} O^{(m)}_{ss''} \int_{-\infty}^{\infty} \nonumber \\ &&\phi^*_{s'}(x) \phi_{s''}(x) dx.
\label{eq-Oint}
\end{eqnarray}
Using the orthonormal property of the functional bases, one has $\int_{-\infty}^{\infty} \phi^*_{s'}(x) \phi_{s''}(x) dx = \delta_{s's''}$, and eventually 
\begin{eqnarray}
\int_{-\infty}^{\infty} \phi^*_{s'}(x) \hat{O}^{(m)}[\phi_{s}(x)] dx 
&=& \sum_{s''=0}^{\mathcal{D}-1} O^{(m)}_{ss''} \delta_{s's''} \nonumber \\ && = O^{(m)}_{ss'}.
\label{eq-Oint1}
\end{eqnarray}

\section{Infinite coupled quantum oscillators}

We consider the infinitely coupled quantum oscillators (iCQOs) in one dimension as an example. The Hamiltonian reads
\begin{eqnarray}
\hat{H}^{\text{HO}} =&& \frac{1}{2} \sum_{n=1}^{N} \left( -\frac{\partial^{2}}{\partial x_{n}^{2}} + \omega_n^2 x_n^2 \right) + \gamma \sum_{m=1}^{N-1} x_m x_{m+1} \nonumber \\ &&+ \tilde{\gamma} \sum_{m=1}^{N-2} x_m x_{m+1} x_{m+2},
\label{eq-HOH}
\end{eqnarray}
where we take the number of oscillators $N \rightarrow \infty$ and the natural frequency $\omega_n=1$. The strengths of the two- and three-body coupling are denoted as $\gamma$ and $\tilde{\gamma}$, respectively. We also take the Planck constant to be $\hbar = 1$ for simplicity.

A natural choice for the orthonormal functional bases is the Gaussian bases
\begin{eqnarray}
	\phi_s(x)=\left(\frac{1}{2^ss!\sqrt{\pi}}\right)^{\frac{1}{2}}e^{-\frac{x^2}{2}}h_s(x),
	\label{eq-HOphi}
\end{eqnarray}
with $h_s(x)$ the Hermitian polynomial. Taking the Gaussian bases and applying Eq.~(\ref{eq-Oint1}) to the partial differentiation operator $\hat{\partial} = \frac{\partial}{\partial x} $, we have 
\begin{equation}
\label{eq-D}
\partial_{s's} = \int_{-\infty}^{\infty} \phi^*_{s'}(x)\hat{\partial}[\phi_s(x)]dx = \left\{
		\begin{array}{lr}
		\sqrt{\frac{s+1}{2}}, & s'=s-1; \\
		-\sqrt{\frac{s+1}{2}}, & s'=s+1. \\
		\end{array}
\right.
\end{equation}

Another example is the operation $\hat{X} = x$. We have 
\begin{equation}
\label{eq-X}
X_{ss'} = \int_{-\infty}^{\infty} \phi^{\ast}_{s}(x) x  \phi_{s'}(x) dx = \left\{
		\begin{array}{lr}
		\sqrt{\frac{s+1}{2}}, & s'=s-1; \\
		\sqrt{\frac{s+1}{2}}, & s'=s+1. \\
		\end{array}
\right.
\end{equation}

The Hamiltonian in Eq.~(\ref{eq-HOphi}) can be expressed in terms of $\hat{\partial}$ and $\hat{X}$. Taking the Hamiltonian without the three-body terms as an example ($\tilde{\gamma}=0$), we have $\hat{H} = \sum_{n} \hat{H}_{n,n+1}$, where the local Hamiltonian satisfies
\begin{eqnarray}
\hat{H}_{n, n+1} =&& -\frac{1}{4}(\hat{\partial}^{2}_{n} + \hat{\partial}^{2}_{n+1} - \hat{X}^{2}_{n} - \hat{X}^{2}_{n+1}) \nonumber \\ && + \gamma \hat{X}_{n} \hat{X}_{n+1},
\label{eq-HOHn}
\end{eqnarray}
with $\hat{\partial}_{n}  = \partial / \partial x_{n}$ and $\hat{X}_{n}  = x_{n}$.

\section{Infinite time-evolving block decimation algorithm}

\begin{figure}[tbp]
	\centering
	\includegraphics[angle=0,width=0.9\linewidth]{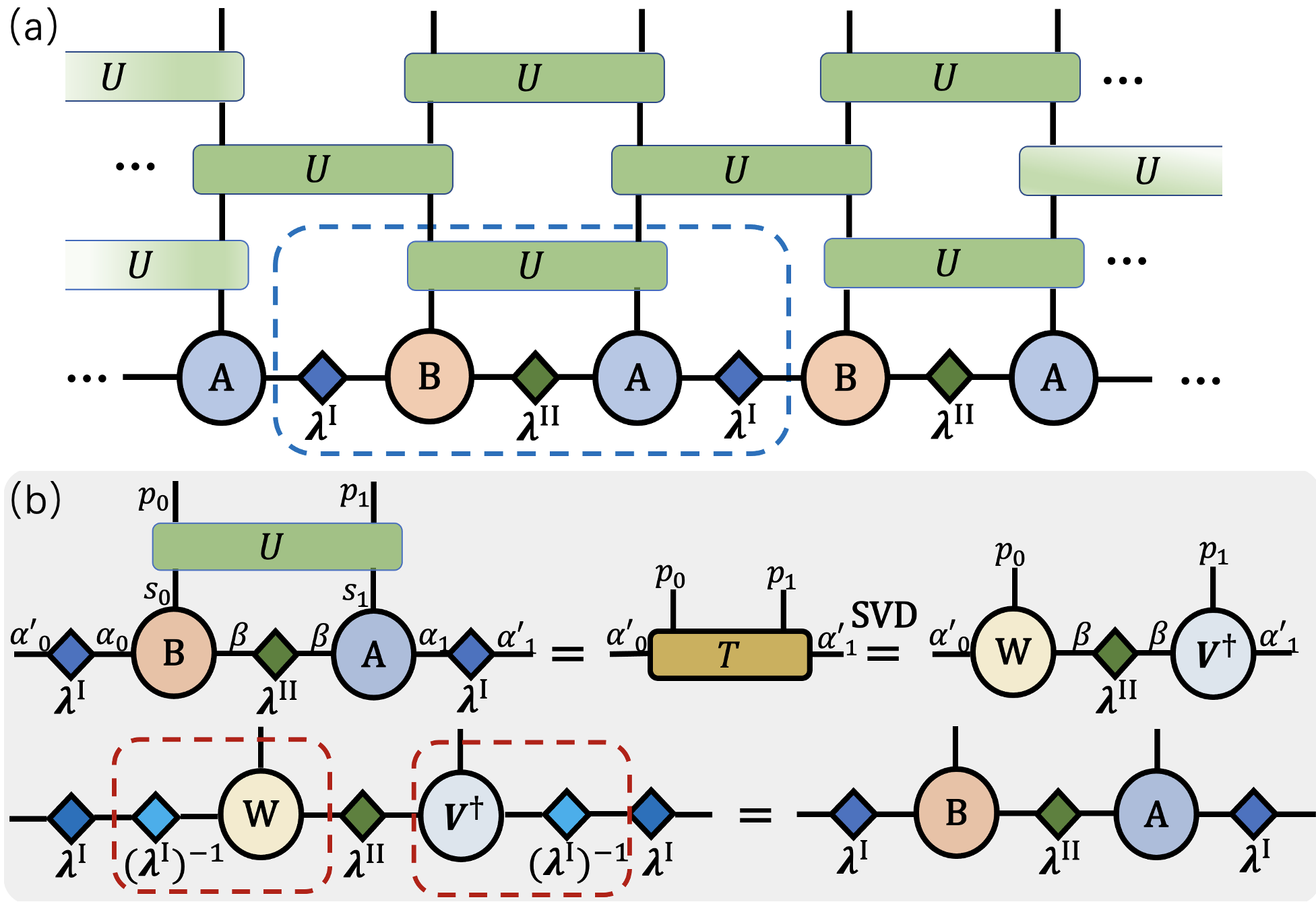}
	\caption{(Color online) (a) An illustration of the imaginary-time evolution in the iTEBD algorithm, where $\hat{U}$ represents the local evolution operator. (b) The tensor contraction for computing the local evolution. See Eqs.~(\ref{eq-T})-(\ref{eq-A}).}
	\label{fig-flow2}
\end{figure}

The ground state of a given Hamiltonian can be simulated using the infinite time-evolving block decimation (iTEBD) algorithm~\cite{itebd1}. We start from the definition of the time-evolution operator 
\begin{equation}
\hat{U}(\tau) = \prod_{n} e^{-\tau \hat{H}_{n,n+1}},
	\label{eq-evolu}
\end{equation}
with $\tau$ a small positive number representing a short time slice. Note that $\hat{U}(\tau)$ is the product of infinitely-many copies of local time-evolution operator $\hat{U} = e^{-\tau \hat{H}_{n,n+1}}$. The ground state $|\psi_{0}\rangle$ can be approached by evolving an initial state $|\psi_{0}\rangle$ until it converges, where we have
\begin{equation}
	\frac{1}{Z}\lim_{K \to \infty} (\prod_{n}\hat{U}(\tau))^{K} = |\psi_{0}\rangle,
	\label{eq-MPS}
\end{equation}
with $Z$ the normalization factor. An illustration of the evolution is given in Fig.~\ref{fig-flow2}(a).

With the presence of three-body terms, we assume the MPS to be 6-tensor translationally invariant, which is formed by infinitely-many copies of $6$ third-order tensors and $6$ spectra (diagonal matrices). The three-body evolution operator can be readily defined to implement the imaginary-time evolution with the presence of three-body coupling terms. Below, we still take the simulation without three-body terms ($\tilde{\gamma}=0$) as an example to explain the algorithm. 

Similar to Eq.~(\ref{eq-MPStensorO}), the local operators (copies of $\hat{U}$) are contracted with the local tensors in the MPS sequentially. For instance, considering the contraction of  $\hat{U}$, $\boldsymbol{A}$, $\boldsymbol{B}$, and the relevant spectra, we have
\begin{equation}
\begin{split}
T_{\alpha'_0 p_0 p_1\alpha_1'} = \sum_{\beta\alpha_0\alpha_1s_0s_1} \lambda_{\alpha'_0 \alpha_0}^{\text{I}} B_{\alpha_0 s_0\beta} \lambda_{\beta \beta}^{\text{II}} A_{\beta s_1\alpha_1}\lambda_{\alpha_1\alpha_1'}^{\text{I}} U_{s_0 s_1 p_0 p_1}
\end{split}
	\label{eq-T}
\end{equation}
with $U_{s_0 s_1 p_0 p_1} = \langle s_0 s_1|\hat{U}| p_0 p_1\rangle$. See Fig.~\ref{fig-flow}(b).

To restore the MPS form before contraction, we perform SVD on $\boldsymbol{T}$ as
\begin{equation}
\begin{split}
\boldsymbol{T}_{[0,1], [2,3]} = \boldsymbol{W}\boldsymbol{\lambda}^{\text{II}}\boldsymbol{V}^\dag
\end{split}
	\label{eq-T01}
\end{equation}
with $\boldsymbol{T}_{[0,1], [2,3]}$ being the matrization of $\boldsymbol{T}$ by reshaping its first two indexes into one index and the rest two indices into another. See the first line of Fig.~\ref{fig-flow}(b). The spectrum $\boldsymbol{\lambda}^{\text{II}}$ is updated by the singular spectrum obtained in the above SVD.

Obviously, the new indexes obtained by SVD and the new spectrum $\boldsymbol{\lambda}^{\text{II}}$ possess a larger dimension. Without any truncation, the dimension will increase exponentially with the evolution time. What we do in the iTEBD algorithm is if this dimension exceeds the preset maximum (denoted as $\chi$), it should be truncated to $\chi$ by keeping only the $\chi$-largest singular values and the corresponding singular vectors. 

Finally, one restores the spectra located at the two sides by inserting $\boldsymbol{\lambda}^{\text{I}} (\boldsymbol{\lambda}^{\text{I}})^{-1}$, as shown in the last line of Fig.~\ref{fig-flow}(b). The two third-order tensors are obtained as
\begin{equation}
A_{\alpha s\beta}=(\mathbf{\lambda}^{\text{I}})^{-1}_{\alpha\alpha}W_{\alpha s\beta}, \ \ 
B_{\alpha s\beta}=V^*_{\alpha s\beta}(\mathbf{\lambda}^{\text{I}})^{-1}_{\beta\beta}.
	\label{eq-A}
\end{equation}

\section{Canonical form of matrix product state}

\begin{figure}[tbp]
	\centering
	\includegraphics[angle=0,width=0.9\linewidth]{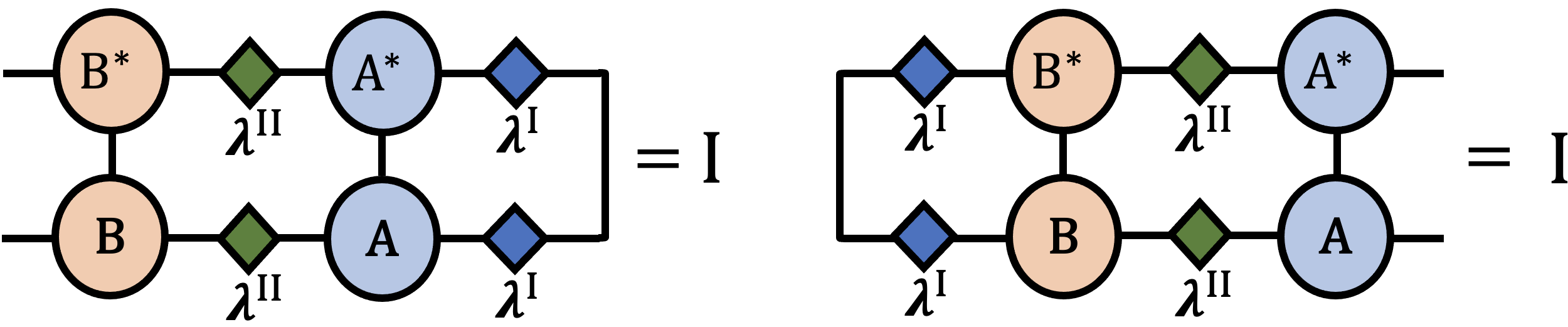}
	\caption{(Color online) The canonical conditions of MPS. See Eq.~(\ref{eq-can}) for the contraction formula corresponding to the left panel.}
	\label{fig-flow1}
\end{figure}

With $\tau \to 0$, the evolution given by $\hat{U}(\tau)$ is close to be identical. Therefore, the evolution drives the MPS to the so-called canonical form~\cite{itebd}, where the tensors satisfy the canonical conditions illustrated in Fig.~\ref{fig-flow1}. As an example, the contraction given by the left panel reads
\begin{align}
&\sum_{s_0s_1\alpha_0\alpha_0'\alpha_1\alpha_1'}   B_{\beta_0 s_0\alpha_0}^* B_{\beta_0's_0 \alpha_0'} \lambda_{\alpha_0 \alpha_0}^{\text{II}} \nonumber \\ &\lambda_{\alpha_0'\alpha_0'}^{\text{II}} A_{\alpha_0 s_1\beta_1}^* A_{\alpha_0' s_1\beta_1} (\lambda_{\beta_{1}\beta_1}^{\text{I}})^{2}  = I_{\beta_{0} \beta_0'},
\label{eq-can}
\end{align}
with $\boldsymbol{I}$ representing the identity. The canonical conditions can be generalized to the cases with $K$-tensor translational invariance.

We here utilize two advantages of the MPS being in the canonical form. First, the truncations implemented in the iTEBD algorithm are (nearly) optimal. By ``nearly'', we mean that the MPS approaches (but is not in) the strict canonical form in the nearly-identical evolution given by $\hat{U}(\tau)$. Second, it has been proven that the spectra ($\boldsymbol{\lambda}^{\text{I}}$ and $\boldsymbol{\lambda}^{\text{II}}$) give the entanglement spectra for the bipartitions at the corresponding positions. These two advantages hold not only for MPS representing the quantum many-body states of lattice models, but also the functional MPS's that represent the wave-functions in the continuous space as well as for the functional MPS's with $K$-tensor translational invariance.

\section{Simulation of local fidelity}

\begin{figure}[tbp]
	\centering
	\includegraphics[angle=0,width=0.9\linewidth]{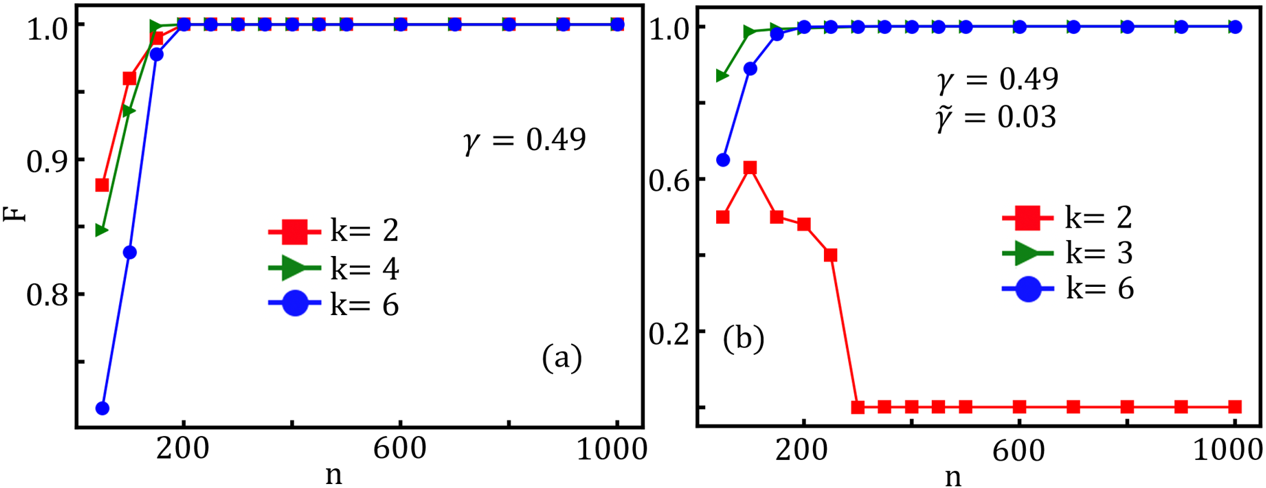}
	\caption{(Color online) Local fidelity $F$ from the fidelity $f = |\langle \psi(K, n)|\psi(K'=6, n'=1000)\rangle|^2$ versus the evolution step $n$ with different translational invariances ($K=2, 3, 4, 6$). $F$ is calculated as the norm square of the largest eigenvalue of the transfer matrix of $f$ (see Fig.~\ref{fig-F2s} for the definition of transfer matrix). \label{fig-F2}}
\end{figure}

\begin{figure}[tbp]
	\centering
	\includegraphics[angle=0,width=0.9\linewidth]{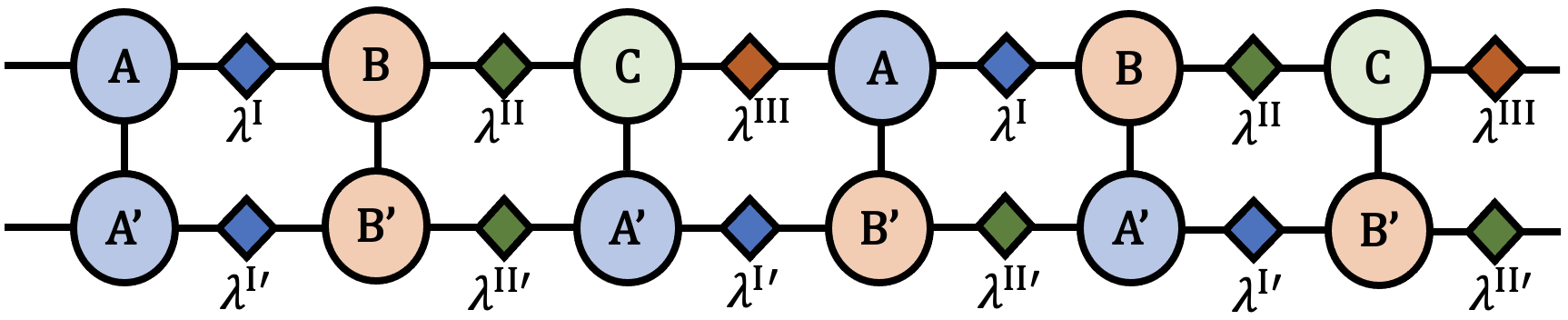}
	\caption{(Color online) The illustration of the transfer matrix for $ |\langle \psi(K=2, n)|\psi(K'=3, n')\rangle|$. We have ignored to illustrate the transpose conjugation of the tensors for simplicity.}
	\label{fig-F2s}
\end{figure}

Fig.~\ref{fig-F2} shows the convergence with the MPS's of different translational symmetries. We consider the $K$-tensor translationally invariant MPS's with  $K = 2, 3, 4, 6$ and show the ``local'' fidelity $F$ that can give the fidelity $f = |\langle \psi(K, n)|\psi(K'=6, n'=1000)\rangle|^2$ versus the number of evolution steps $n$. Note $|\psi(K, n)\rangle$ denotes a $K$-tensor translationally invariant MPS after $n$ evolution steps.

The local fidelity is calculated by the largest eigenvalue of the transfer matrix of $|\langle \psi(K, n)|\psi(K', n')\rangle|$. See Fig.~\ref{fig-F2s} as the illustration of the transfer matrix with $K=2$ and $K'=3$ as an example. One should take $\tilde{K}=6$ tensors from each of the matrix to define the transfer matrix, with $\tilde{K}$ the least common multiple of $K$ and $K'$. The local fidelity is calculated as the norm square of the largest eigenvalue of the transfer matrix. Specifically, we have
\begin{equation}
f = F^{\frac{N}{\tilde{K}}},
	\label{eq-Ff}
\end{equation}
with the system size $N \to \infty$ in the thermodynamic limit. One always has $f\to 0$ when the difference between the local tensors of two MPS's lead the dominant eigenvalue to be smaller than 1. Therefore, we choose the local fidelity $F$ to characterize the similarity between two infinite-size MPS's rather than the fidelity $f$.

In Fig.~\ref{fig-F2}(a), we take $\gamma=0.49$ and $\tilde{\gamma}=0$, where the Hamiltonian consists of just two-body couplings. All local fidelities converge to $1$, meaning the MPS's stably converge to $|\psi(K=6, n=1000)\rangle$. In Fig.~\ref{fig-F2}(b), we introduce the three-body couplings by taking $\tilde{\gamma}=0.03$. The MPS with $6$-tensor translational invariance converges well as $n$ increases. However, the fidelity $F = |\langle \psi(K=2, n)|\psi(K'=6, n'=1000)\rangle|^2$ converges to about $0.1$. This is due to the inconsistency between the translational invariance ($K=2$) and the 3-body coupling terns. Therefore, we trust the results with $K=6$.

\input{iFTN.bbl}
\end{document}

%% file: iFTN.bbl
%

%% file: iFTN.bbl
\begin{thebibliography}{44}%
\makeatletter
\providecommand \@ifxundefined [1]{%
 \@ifx{#1\undefined}
}%
\providecommand \@ifnum [1]{%
 \ifnum #1\expandafter \@firstoftwo
 \else \expandafter \@secondoftwo
 \fi
}%
\providecommand \@ifx [1]{%
 \ifx #1\expandafter \@firstoftwo
 \else \expandafter \@secondoftwo
 \fi
}%
\providecommand \natexlab [1]{#1}%
\providecommand \enquote  [1]{``#1''}%
\providecommand \bibnamefont  [1]{#1}%
\providecommand \bibfnamefont [1]{#1}%
\providecommand \citenamefont [1]{#1}%
\providecommand \href@noop [0]{\@secondoftwo}%
\providecommand \href [0]{\begingroup \@sanitize@url \@href}%
\providecommand \@href[1]{\@@startlink{#1}\@@href}%
\providecommand \@@href[1]{\endgroup#1\@@endlink}%
\providecommand \@sanitize@url [0]{\catcode `\\12\catcode `\$12\catcode
  `\&12\catcode `\#12\catcode `\^12\catcode `\_12\catcode `\%12\relax}%
\providecommand \@@startlink[1]{}%
\providecommand \@@endlink[0]{}%
\providecommand \url  [0]{\begingroup\@sanitize@url \@url }%
\providecommand \@url [1]{\endgroup\@href {#1}{\urlprefix }}%
\providecommand \urlprefix  [0]{URL }%
\providecommand \Eprint [0]{\href }%
\providecommand \doibase [0]{http://dx.doi.org/}%
\providecommand \selectlanguage [0]{\@gobble}%
\providecommand \bibinfo  [0]{\@secondoftwo}%
\providecommand \bibfield  [0]{\@secondoftwo}%
\providecommand \translation [1]{[#1]}%
\providecommand \BibitemOpen [0]{}%
\providecommand \bibitemStop [0]{}%
\providecommand \bibitemNoStop [0]{.\EOS\space}%
\providecommand \EOS [0]{\spacefactor3000\relax}%
\providecommand \BibitemShut  [1]{\csname bibitem#1\endcsname}%
\let\auto@bib@innerbib\@empty
\bibitem [{\citenamefont {F.~Verstraete}\ and\ \citenamefont
  {Cirac}(2008)}]{TNs3}%
  \BibitemOpen
  \bibfield  {author} {\bibinfo {author} {\bibfnamefont {V.~Murg}\ \bibnamefont
  {F.~Verstraete}}\ and\ \bibinfo {author} {\bibfnamefont {J.I.}\ \bibnamefont
  {Cirac}},\ }\bibfield  {title} {\enquote {\bibinfo {title} {Matrix product
  states, projected entangled pair states, and variational renormalization
  group methods for quantum spin systems},}\ }\href {\doibase
  10.1080/14789940801912366} {\bibfield  {journal} {\bibinfo  {journal}
  {Advances in Physics}\ }\textbf {\bibinfo {volume} {57}},\ \bibinfo {pages}
  {143--224} (\bibinfo {year} {2008})}\BibitemShut {NoStop}%
\bibitem [{\citenamefont {Cirac}\ and\ \citenamefont
  {Verstraete}(2009)}]{TNs1}%
  \BibitemOpen
  \bibfield  {author} {\bibinfo {author} {\bibfnamefont {J~Ignacio}\
  \bibnamefont {Cirac}}\ and\ \bibinfo {author} {\bibfnamefont {Frank}\
  \bibnamefont {Verstraete}},\ }\bibfield  {title} {\enquote {\bibinfo {title}
  {Renormalization and tensor product states in spin chains and lattices},}\
  }\href {\doibase 10.1088/1751-8113/42/50/504004} {\bibfield  {journal}
  {\bibinfo  {journal} {Journal of Physics A: Mathematical and Theoretical}\
  }\textbf {\bibinfo {volume} {42}},\ \bibinfo {pages} {504004} (\bibinfo
  {year} {2009})}\BibitemShut {NoStop}%
\bibitem [{\citenamefont {Orús}(2014)}]{TNs2}%
  \BibitemOpen
  \bibfield  {author} {\bibinfo {author} {\bibfnamefont {Román}\ \bibnamefont
  {Orús}},\ }\bibfield  {title} {\enquote {\bibinfo {title} {A practical
  introduction to tensor networks: Matrix product states and projected
  entangled pair states},}\ }\href {\doibase
  https://doi.org/10.1016/j.aop.2014.06.013} {\bibfield  {journal} {\bibinfo
  {journal} {Annals of Physics}\ }\textbf {\bibinfo {volume} {349}},\ \bibinfo
  {pages} {117--158} (\bibinfo {year} {2014})}\BibitemShut {NoStop}%
\bibitem [{\citenamefont {Ran}\ \emph {et~al.}()\citenamefont {Ran},
  \citenamefont {Tirrito}, \citenamefont {Peng}, \citenamefont {Chen},
  \citenamefont {Tagliacozzo}, \citenamefont {Su},\ and\ \citenamefont
  {Lewenstein}}]{TNbook}%
  \BibitemOpen
  \bibfield  {author} {\bibinfo {author} {\bibfnamefont {Shi-Ju}\ \bibnamefont
  {Ran}}, \bibinfo {author} {\bibfnamefont {Emanuele}\ \bibnamefont {Tirrito}},
  \bibinfo {author} {\bibfnamefont {Cheng}\ \bibnamefont {Peng}}, \bibinfo
  {author} {\bibfnamefont {Xi}~\bibnamefont {Chen}}, \bibinfo {author}
  {\bibfnamefont {Luca}\ \bibnamefont {Tagliacozzo}}, \bibinfo {author}
  {\bibfnamefont {Gang}\ \bibnamefont {Su}}, \ and\ \bibinfo {author}
  {\bibfnamefont {Maciej}\ \bibnamefont {Lewenstein}},\ }\href@noop {} {}\
  (\bibinfo  {publisher} {Springer, Cham})\BibitemShut {NoStop}%
\bibitem [{\citenamefont {WEN}(1990)}]{toporder0}%
  \BibitemOpen
  \bibfield  {author} {\bibinfo {author} {\bibfnamefont {X.~G.}\ \bibnamefont
  {WEN}},\ }\bibfield  {title} {\enquote {\bibinfo {title} {Topological orders
  in rigid states},}\ }\href {\doibase 10.1142/S0217979290000139} {\bibfield
  {journal} {\bibinfo  {journal} {International Journal of Modern Physics B}\
  }\textbf {\bibinfo {volume} {04}},\ \bibinfo {pages} {239--271} (\bibinfo
  {year} {1990})}\BibitemShut {NoStop}%
\bibitem [{\citenamefont {Levin}\ and\ \citenamefont {Wen}(2006)}]{toporder1}%
  \BibitemOpen
  \bibfield  {author} {\bibinfo {author} {\bibfnamefont {Michael}\ \bibnamefont
  {Levin}}\ and\ \bibinfo {author} {\bibfnamefont {Xiao-Gang}\ \bibnamefont
  {Wen}},\ }\bibfield  {title} {\enquote {\bibinfo {title} {Detecting
  topological order in a ground state wave function},}\ }\href {\doibase
  10.1103/PhysRevLett.96.110405} {\bibfield  {journal} {\bibinfo  {journal}
  {Phys. Rev. Lett.}\ }\textbf {\bibinfo {volume} {96}},\ \bibinfo {pages}
  {110405} (\bibinfo {year} {2006})}\BibitemShut {NoStop}%
\bibitem [{\citenamefont {Kitaev}\ and\ \citenamefont
  {Preskill}(2006)}]{toporder2}%
  \BibitemOpen
  \bibfield  {author} {\bibinfo {author} {\bibfnamefont {Alexei}\ \bibnamefont
  {Kitaev}}\ and\ \bibinfo {author} {\bibfnamefont {John}\ \bibnamefont
  {Preskill}},\ }\bibfield  {title} {\enquote {\bibinfo {title} {Topological
  entanglement entropy},}\ }\href {\doibase 10.1103/PhysRevLett.96.110404}
  {\bibfield  {journal} {\bibinfo  {journal} {Physical Review Letters}\
  }\textbf {\bibinfo {volume} {96}},\ \bibinfo {pages} {110404} (\bibinfo
  {year} {2006})}\BibitemShut {NoStop}%
\bibitem [{\citenamefont {Jiang}(2012)}]{toporder3}%
  \BibitemOpen
  \bibfield  {author} {\bibinfo {author} {\bibfnamefont {Wang Z. Balents~L.}\
  \bibnamefont {Jiang}, \bibfnamefont {HC.}},\ }\bibfield  {title} {\enquote
  {\bibinfo {title} {Identifying topological order by entanglement entropy},}\
  }\href {\doibase 10.1038/nphys2465} {\bibfield  {journal} {\bibinfo
  {journal} {Nature Phys}\ }\textbf {\bibinfo {volume} {8}},\ \bibinfo {pages}
  {902--905} (\bibinfo {year} {2012})}\BibitemShut {NoStop}%
\bibitem [{\citenamefont {Evenbly}\ and\ \citenamefont
  {Vidal}(2015)}]{TNCriticality1}%
  \BibitemOpen
  \bibfield  {author} {\bibinfo {author} {\bibfnamefont {G.}~\bibnamefont
  {Evenbly}}\ and\ \bibinfo {author} {\bibfnamefont {G.}~\bibnamefont
  {Vidal}},\ }\bibfield  {title} {\enquote {\bibinfo {title} {Tensor network
  renormalization},}\ }\href {\doibase 10.1103/PhysRevLett.115.180405}
  {\bibfield  {journal} {\bibinfo  {journal} {Physical Review Letters}\
  }\textbf {\bibinfo {volume} {115}},\ \bibinfo {pages} {180405} (\bibinfo
  {year} {2015})}\BibitemShut {NoStop}%
\bibitem [{\citenamefont {Yang}\ \emph {et~al.}(2023)\citenamefont {Yang},
  \citenamefont {Vanhecke},\ and\ \citenamefont {Schuch}}]{TNCriticality2}%
  \BibitemOpen
  \bibfield  {author} {\bibinfo {author} {\bibfnamefont {Mingru}\ \bibnamefont
  {Yang}}, \bibinfo {author} {\bibfnamefont {Bram}\ \bibnamefont {Vanhecke}}, \
  and\ \bibinfo {author} {\bibfnamefont {Norbert}\ \bibnamefont {Schuch}},\
  }\bibfield  {title} {\enquote {\bibinfo {title} {Detecting emergent
  continuous symmetries at quantum criticality},}\ }\href {\doibase
  10.1103/PhysRevLett.131.036505} {\bibfield  {journal} {\bibinfo  {journal}
  {Physical Review Letters}\ }\textbf {\bibinfo {volume} {131}},\ \bibinfo
  {pages} {036505} (\bibinfo {year} {2023})}\BibitemShut {NoStop}%
\bibitem [{\citenamefont {Vidal}\ \emph {et~al.}(2003)\citenamefont {Vidal},
  \citenamefont {Latorre}, \citenamefont {Rico},\ and\ \citenamefont
  {Kitaev}}]{entropy}%
  \BibitemOpen
  \bibfield  {author} {\bibinfo {author} {\bibfnamefont {G.}~\bibnamefont
  {Vidal}}, \bibinfo {author} {\bibfnamefont {J.~I.}\ \bibnamefont {Latorre}},
  \bibinfo {author} {\bibfnamefont {E.}~\bibnamefont {Rico}}, \ and\ \bibinfo
  {author} {\bibfnamefont {A.}~\bibnamefont {Kitaev}},\ }\bibfield  {title}
  {\enquote {\bibinfo {title} {Entanglement in quantum critical phenomena},}\
  }\href {\doibase 10.1103/PhysRevLett.90.227902} {\bibfield  {journal}
  {\bibinfo  {journal} {Phys. Rev. Lett.}\ }\textbf {\bibinfo {volume} {90}},\
  \bibinfo {pages} {227902} (\bibinfo {year} {2003})}\BibitemShut {NoStop}%
\bibitem [{\citenamefont {Tagliacozzo}\ \emph {et~al.}(2008)\citenamefont
  {Tagliacozzo}, \citenamefont {de~Oliveira}, \citenamefont {Iblisdir},\ and\
  \citenamefont {Latorre}}]{mps}%
  \BibitemOpen
  \bibfield  {author} {\bibinfo {author} {\bibfnamefont {L.}~\bibnamefont
  {Tagliacozzo}}, \bibinfo {author} {\bibfnamefont {Thiago.~R.}\ \bibnamefont
  {de~Oliveira}}, \bibinfo {author} {\bibfnamefont {S.}~\bibnamefont
  {Iblisdir}}, \ and\ \bibinfo {author} {\bibfnamefont {J.~I.}\ \bibnamefont
  {Latorre}},\ }\bibfield  {title} {\enquote {\bibinfo {title} {Scaling of
  entanglement support for matrix product states},}\ }\href {\doibase
  10.1103/PhysRevB.78.024410} {\bibfield  {journal} {\bibinfo  {journal} {Phys.
  Rev. B}\ }\textbf {\bibinfo {volume} {78}},\ \bibinfo {pages} {024410}
  (\bibinfo {year} {2008})}\BibitemShut {NoStop}%
\bibitem [{\citenamefont {Pollmann}\ \emph {et~al.}(2009)\citenamefont
  {Pollmann}, \citenamefont {Mukerjee}, \citenamefont {Turner},\ and\
  \citenamefont {Moore}}]{Scaling}%
  \BibitemOpen
  \bibfield  {author} {\bibinfo {author} {\bibfnamefont {Frank}\ \bibnamefont
  {Pollmann}}, \bibinfo {author} {\bibfnamefont {Subroto}\ \bibnamefont
  {Mukerjee}}, \bibinfo {author} {\bibfnamefont {Ari~M.}\ \bibnamefont
  {Turner}}, \ and\ \bibinfo {author} {\bibfnamefont {Joel~E.}\ \bibnamefont
  {Moore}},\ }\bibfield  {title} {\enquote {\bibinfo {title} {Theory of
  finite-entanglement scaling at one-dimensional quantum critical points},}\
  }\href {\doibase 10.1103/PhysRevLett.102.255701} {\bibfield  {journal}
  {\bibinfo  {journal} {Physical Review Letters}\ }\textbf {\bibinfo {volume}
  {102}},\ \bibinfo {pages} {255701} (\bibinfo {year} {2009})}\BibitemShut
  {NoStop}%
\bibitem [{\citenamefont {Ge}\ and\ \citenamefont {Eisert}(2016)}]{arealaw2}%
  \BibitemOpen
  \bibfield  {author} {\bibinfo {author} {\bibfnamefont {Yimin}\ \bibnamefont
  {Ge}}\ and\ \bibinfo {author} {\bibfnamefont {Jens}\ \bibnamefont {Eisert}},\
  }\bibfield  {title} {\enquote {\bibinfo {title} {Area laws and efficient
  descriptions of quantum many-body states},}\ }\href {\doibase
  10.1088/1367-2630/18/8/083026} {\bibfield  {journal} {\bibinfo  {journal}
  {New Journal of Physics}\ }\textbf {\bibinfo {volume} {18}},\ \bibinfo
  {pages} {083026} (\bibinfo {year} {2016})}\BibitemShut {NoStop}%
\bibitem [{\citenamefont {Eisert}\ \emph {et~al.}(2010)\citenamefont {Eisert},
  \citenamefont {Cramer},\ and\ \citenamefont {Plenio}}]{arealaw1}%
  \BibitemOpen
  \bibfield  {author} {\bibinfo {author} {\bibfnamefont {J.}~\bibnamefont
  {Eisert}}, \bibinfo {author} {\bibfnamefont {M.}~\bibnamefont {Cramer}}, \
  and\ \bibinfo {author} {\bibfnamefont {M.~B.}\ \bibnamefont {Plenio}},\
  }\bibfield  {title} {\enquote {\bibinfo {title} {Colloquium: Area laws for
  the entanglement entropy},}\ }\href {\doibase 10.1103/RevModPhys.82.277}
  {\bibfield  {journal} {\bibinfo  {journal} {Rev. Mod. Phys.}\ }\textbf
  {\bibinfo {volume} {82}},\ \bibinfo {pages} {277--306} (\bibinfo {year}
  {2010})}\BibitemShut {NoStop}%
\bibitem [{\citenamefont {Zou}\ \emph {et~al.}(2018)\citenamefont {Zou},
  \citenamefont {Milsted},\ and\ \citenamefont {Vidal}}]{critical1}%
  \BibitemOpen
  \bibfield  {author} {\bibinfo {author} {\bibfnamefont {Yijian}\ \bibnamefont
  {Zou}}, \bibinfo {author} {\bibfnamefont {Ashley}\ \bibnamefont {Milsted}}, \
  and\ \bibinfo {author} {\bibfnamefont {Guifre}\ \bibnamefont {Vidal}},\
  }\bibfield  {title} {\enquote {\bibinfo {title} {Conformal data and
  renormalization group flow in critical quantum spin chains using periodic
  uniform matrix product states},}\ }\href {\doibase
  10.1103/PhysRevLett.121.230402} {\bibfield  {journal} {\bibinfo  {journal}
  {Physical Review Letters}\ }\textbf {\bibinfo {volume} {121}},\ \bibinfo
  {pages} {230402} (\bibinfo {year} {2018})}\BibitemShut {NoStop}%
\bibitem [{\citenamefont {Huang}\ \emph {et~al.}(2024)\citenamefont {Huang},
  \citenamefont {Zhang}, \citenamefont {L\"auchli}, \citenamefont {Haegeman},
  \citenamefont {Verstraete},\ and\ \citenamefont
  {Vanderstraeten}}]{mps_scaling}%
  \BibitemOpen
  \bibfield  {author} {\bibinfo {author} {\bibfnamefont {Rui-Zhen}\
  \bibnamefont {Huang}}, \bibinfo {author} {\bibfnamefont {Long}\ \bibnamefont
  {Zhang}}, \bibinfo {author} {\bibfnamefont {Andreas~M.}\ \bibnamefont
  {L\"auchli}}, \bibinfo {author} {\bibfnamefont {Jutho}\ \bibnamefont
  {Haegeman}}, \bibinfo {author} {\bibfnamefont {Frank}\ \bibnamefont
  {Verstraete}}, \ and\ \bibinfo {author} {\bibfnamefont {Laurens}\
  \bibnamefont {Vanderstraeten}},\ }\bibfield  {title} {\enquote {\bibinfo
  {title} {Emergent conformal boundaries from finite-entanglement scaling in
  matrix product states},}\ }\href {\doibase 10.1103/PhysRevLett.132.086503}
  {\bibfield  {journal} {\bibinfo  {journal} {Physical Review Letters}\
  }\textbf {\bibinfo {volume} {132}},\ \bibinfo {pages} {086503} (\bibinfo
  {year} {2024})}\BibitemShut {NoStop}%
\bibitem [{\citenamefont {\"Ostlund}\ and\ \citenamefont
  {Rommer}(1995)}]{mpschushi1}%
  \BibitemOpen
  \bibfield  {author} {\bibinfo {author} {\bibfnamefont {Stellan}\ \bibnamefont
  {\"Ostlund}}\ and\ \bibinfo {author} {\bibfnamefont {Stefan}\ \bibnamefont
  {Rommer}},\ }\bibfield  {title} {\enquote {\bibinfo {title} {Thermodynamic
  limit of density matrix renormalization},}\ }\href {\doibase
  10.1103/PhysRevLett.75.3537} {\bibfield  {journal} {\bibinfo  {journal}
  {Physical Review Letters}\ }\textbf {\bibinfo {volume} {75}},\ \bibinfo
  {pages} {3537--3540} (\bibinfo {year} {1995})}\BibitemShut {NoStop}%
\bibitem [{\citenamefont {Rommer}\ and\ \citenamefont
  {\"Ostlund}(1997)}]{mpschushi2}%
  \BibitemOpen
  \bibfield  {author} {\bibinfo {author} {\bibfnamefont {Stefan}\ \bibnamefont
  {Rommer}}\ and\ \bibinfo {author} {\bibfnamefont {Stellan}\ \bibnamefont
  {\"Ostlund}},\ }\bibfield  {title} {\enquote {\bibinfo {title} {Class of
  ansatz wave functions for one-dimensional spin systems and their relation to
  the density matrix renormalization group},}\ }\href {\doibase
  10.1103/PhysRevB.55.2164} {\bibfield  {journal} {\bibinfo  {journal}
  {Physical Review B}\ }\textbf {\bibinfo {volume} {55}},\ \bibinfo {pages}
  {2164--2181} (\bibinfo {year} {1997})}\BibitemShut {NoStop}%
\bibitem [{\citenamefont {Verstraete}\ and\ \citenamefont
  {Cirac}(2006)}]{TNmps}%
  \BibitemOpen
  \bibfield  {author} {\bibinfo {author} {\bibfnamefont {F.}~\bibnamefont
  {Verstraete}}\ and\ \bibinfo {author} {\bibfnamefont {J.~I.}\ \bibnamefont
  {Cirac}},\ }\bibfield  {title} {\enquote {\bibinfo {title} {Matrix product
  states represent ground states faithfully},}\ }\href {\doibase
  10.1103/PhysRevB.73.094423} {\bibfield  {journal} {\bibinfo  {journal}
  {Physical Review B}\ }\textbf {\bibinfo {volume} {73}},\ \bibinfo {pages}
  {094423} (\bibinfo {year} {2006})}\BibitemShut {NoStop}%
\bibitem [{\citenamefont {Hastings}(2006)}]{gapped1d}%
  \BibitemOpen
  \bibfield  {author} {\bibinfo {author} {\bibfnamefont {M.~B.}\ \bibnamefont
  {Hastings}},\ }\bibfield  {title} {\enquote {\bibinfo {title} {Solving gapped
  hamiltonians locally},}\ }\href {\doibase 10.1103/PhysRevB.73.085115}
  {\bibfield  {journal} {\bibinfo  {journal} {Phys. Rev. B}\ }\textbf {\bibinfo
  {volume} {73}},\ \bibinfo {pages} {085115} (\bibinfo {year}
  {2006})}\BibitemShut {NoStop}%
\bibitem [{\citenamefont {King}(1951)}]{mc2}%
  \BibitemOpen
  \bibfield  {author} {\bibinfo {author} {\bibfnamefont {Gilbert~W.}\
  \bibnamefont {King}},\ }\bibfield  {title} {\enquote {\bibinfo {title}
  {Monte-carlo method for solving diffusion problems},}\ }\href {\doibase
  10.1021/ie50503a021} {\bibfield  {journal} {\bibinfo  {journal} {Industrial
  and Engineering Chemistry}\ }\textbf {\bibinfo {volume} {43}},\ \bibinfo
  {pages} {2475--2478} (\bibinfo {year} {1951})}\BibitemShut {NoStop}%
\bibitem [{\citenamefont {Ceperley}\ and\ \citenamefont {Alder}(1986)}]{qmk0}%
  \BibitemOpen
  \bibfield  {author} {\bibinfo {author} {\bibfnamefont {David}\ \bibnamefont
  {Ceperley}}\ and\ \bibinfo {author} {\bibfnamefont {Berni}\ \bibnamefont
  {Alder}},\ }\bibfield  {title} {\enquote {\bibinfo {title} {{Quantum Monte
  Carlo}},}\ }\href {\doibase 10.1126/science.231.4738.555} {\bibfield
  {journal} {\bibinfo  {journal} {Science}\ }\textbf {\bibinfo {volume}
  {231}},\ \bibinfo {pages} {555--560} (\bibinfo {year} {1986})}\BibitemShut
  {NoStop}%
\bibitem [{\citenamefont {Foulkes}\ \emph {et~al.}(2001)\citenamefont
  {Foulkes}, \citenamefont {Mitas}, \citenamefont {Needs},\ and\ \citenamefont
  {Rajagopal}}]{mk2}%
  \BibitemOpen
  \bibfield  {author} {\bibinfo {author} {\bibfnamefont {W.~M.~C.}\
  \bibnamefont {Foulkes}}, \bibinfo {author} {\bibfnamefont {L.}~\bibnamefont
  {Mitas}}, \bibinfo {author} {\bibfnamefont {R.~J.}\ \bibnamefont {Needs}}, \
  and\ \bibinfo {author} {\bibfnamefont {G.}~\bibnamefont {Rajagopal}},\
  }\bibfield  {title} {\enquote {\bibinfo {title} {Quantum monte carlo
  simulations of solids},}\ }\href {\doibase 10.1103/RevModPhys.73.33}
  {\bibfield  {journal} {\bibinfo  {journal} {Reviews of Modern Physics}\
  }\textbf {\bibinfo {volume} {73}},\ \bibinfo {pages} {33--83} (\bibinfo
  {year} {2001})}\BibitemShut {NoStop}%
\bibitem [{\citenamefont {Huggins}(2022)}]{qmk}%
  \BibitemOpen
  \bibfield  {author} {\bibinfo {author} {\bibfnamefont {O’Gorman B.A. Rubin
  N.C. et~al.}\ \bibnamefont {Huggins}, \bibfnamefont {W.J.}},\ }\bibfield
  {title} {\enquote {\bibinfo {title} {Unbiasing fermionic quantum monte carlo
  with a quantum computer},}\ }\href {\doibase 10.1038/s41586-021-04351-z}
  {\bibfield  {journal} {\bibinfo  {journal} {Nature}\ }\textbf {\bibinfo
  {volume} {603}},\ \bibinfo {pages} {416--420} (\bibinfo {year}
  {2022})}\BibitemShut {NoStop}%
\bibitem [{\citenamefont {Sirignano}\ and\ \citenamefont
  {Spiliopoulos}(2018)}]{NN4}%
  \BibitemOpen
  \bibfield  {author} {\bibinfo {author} {\bibfnamefont {Justin}\ \bibnamefont
  {Sirignano}}\ and\ \bibinfo {author} {\bibfnamefont {Konstantinos}\
  \bibnamefont {Spiliopoulos}},\ }\bibfield  {title} {\enquote {\bibinfo
  {title} {Dgm: A deep learning algorithm for solving partial differential
  equations},}\ }\href {\doibase 10.1016/j.jcp.2018.08.029} {\bibfield
  {journal} {\bibinfo  {journal} {Journal of Computational Physics}\ }\textbf
  {\bibinfo {volume} {375}},\ \bibinfo {pages} {1339--1364} (\bibinfo {year}
  {2018})}\BibitemShut {NoStop}%
\bibitem [{\citenamefont {Raissi}\ \emph {et~al.}(2019)\citenamefont {Raissi},
  \citenamefont {Perdikaris},\ and\ \citenamefont {Karniadakis}}]{NN3}%
  \BibitemOpen
  \bibfield  {author} {\bibinfo {author} {\bibfnamefont {M.}~\bibnamefont
  {Raissi}}, \bibinfo {author} {\bibfnamefont {P.}~\bibnamefont {Perdikaris}},
  \ and\ \bibinfo {author} {\bibfnamefont {G.E.}\ \bibnamefont {Karniadakis}},\
  }\bibfield  {title} {\enquote {\bibinfo {title} {Physics-informed neural
  networks: A deep learning framework for solving forward and inverse problems
  involving nonlinear partial differential equations},}\ }\href {\doibase
  https://doi.org/10.1016/j.jcp.2018.10.045} {\bibfield  {journal} {\bibinfo
  {journal} {Journal of Computational Physics}\ }\textbf {\bibinfo {volume}
  {378}},\ \bibinfo {pages} {686--707} (\bibinfo {year} {2019})}\BibitemShut
  {NoStop}%
\bibitem [{\citenamefont {Pfau}\ \emph {et~al.}(2020)\citenamefont {Pfau},
  \citenamefont {Spencer}, \citenamefont {Matthews},\ and\ \citenamefont
  {Foulkes}}]{NN1}%
  \BibitemOpen
  \bibfield  {author} {\bibinfo {author} {\bibfnamefont {David}\ \bibnamefont
  {Pfau}}, \bibinfo {author} {\bibfnamefont {James~S.}\ \bibnamefont
  {Spencer}}, \bibinfo {author} {\bibfnamefont {Alexander G. D.~G.}\
  \bibnamefont {Matthews}}, \ and\ \bibinfo {author} {\bibfnamefont {W.~M.~C.}\
  \bibnamefont {Foulkes}},\ }\bibfield  {title} {\enquote {\bibinfo {title} {Ab
  initio solution of the many-electron schr\"odinger equation with deep neural
  networks},}\ }\href {\doibase 10.1103/PhysRevResearch.2.033429} {\bibfield
  {journal} {\bibinfo  {journal} {Physical Review Research}\ }\textbf {\bibinfo
  {volume} {2}},\ \bibinfo {pages} {033429} (\bibinfo {year}
  {2020})}\BibitemShut {NoStop}%
\bibitem [{\citenamefont {Hermann}\ and\ \citenamefont {No\`e}(2020)}]{NNa}%
  \BibitemOpen
  \bibfield  {author} {\bibinfo {author} {\bibfnamefont {Sch\"atzle~Z.}\
  \bibnamefont {Hermann}, \bibfnamefont {J.}}\ and\ \bibinfo {author}
  {\bibfnamefont {F}~\bibnamefont {No\`e}},\ }\bibfield  {title} {\enquote
  {\bibinfo {title} {Deep-neural-network solution of the electronic
  schr\"odinger equation},}\ }\href {\doibase 10.1038/s41557-020-0544-y}
  {\bibfield  {journal} {\bibinfo  {journal} {Nature Chemistry}\ }\textbf
  {\bibinfo {volume} {12}},\ \bibinfo {pages} {891--897} (\bibinfo {year}
  {2020})}\BibitemShut {NoStop}%
\bibitem [{\citenamefont {cherbela}(2022)}]{NNb}%
  \BibitemOpen
  \bibfield  {author} {\bibinfo {author} {\bibfnamefont {Reisenhofer R. Gerard
  L. et~al.}\ \bibnamefont {cherbela}, \bibfnamefont {M.}},\ }\bibfield
  {title} {\enquote {\bibinfo {title} {Solving the electronic schrödinger
  equation for multiple nuclear geometries with weight-sharing deep neural
  networks},}\ }\href {\doibase 10.1038/s43588-022-00228-x} {\bibfield
  {journal} {\bibinfo  {journal} {Nature Computational Science}\ }\textbf
  {\bibinfo {volume} {2}},\ \bibinfo {pages} {331--341} (\bibinfo {year}
  {2022})}\BibitemShut {NoStop}%
\bibitem [{\citenamefont {Or\'us}\ and\ \citenamefont {Vidal}(2008)}]{itebd}%
  \BibitemOpen
  \bibfield  {author} {\bibinfo {author} {\bibfnamefont {R.}~\bibnamefont
  {Or\'us}}\ and\ \bibinfo {author} {\bibfnamefont {G.}~\bibnamefont {Vidal}},\
  }\bibfield  {title} {\enquote {\bibinfo {title} {Infinite time-evolving block
  decimation algorithm beyond unitary evolution},}\ }\href {\doibase
  10.1103/PhysRevB.78.155117} {\bibfield  {journal} {\bibinfo  {journal}
  {Physical Review B}\ }\textbf {\bibinfo {volume} {78}},\ \bibinfo {pages}
  {155117} (\bibinfo {year} {2008})}\BibitemShut {NoStop}%
\bibitem [{\citenamefont {Vidal}(2007)}]{itebd1}%
  \BibitemOpen
  \bibfield  {author} {\bibinfo {author} {\bibfnamefont {G.}~\bibnamefont
  {Vidal}},\ }\bibfield  {title} {\enquote {\bibinfo {title} {Classical
  simulation of infinite-size quantum lattice systems in one spatial
  dimension},}\ }\href {\doibase 10.1103/PhysRevLett.98.070201} {\bibfield
  {journal} {\bibinfo  {journal} {Physical Review Letters}\ }\textbf {\bibinfo
  {volume} {98}},\ \bibinfo {pages} {070201} (\bibinfo {year}
  {2007})}\BibitemShut {NoStop}%
\bibitem [{\citenamefont {Iblisdir}\ \emph {et~al.}(2007)\citenamefont
  {Iblisdir}, \citenamefont {Or\'us},\ and\ \citenamefont {Latorre}}]{mpscon}%
  \BibitemOpen
  \bibfield  {author} {\bibinfo {author} {\bibfnamefont {S.}~\bibnamefont
  {Iblisdir}}, \bibinfo {author} {\bibfnamefont {R.}~\bibnamefont {Or\'us}}, \
  and\ \bibinfo {author} {\bibfnamefont {J.~I.}\ \bibnamefont {Latorre}},\
  }\bibfield  {title} {\enquote {\bibinfo {title} {Matrix product states
  algorithms and continuous systems},}\ }\href {\doibase
  10.1103/PhysRevB.75.104305} {\bibfield  {journal} {\bibinfo  {journal}
  {Physical Review B}\ }\textbf {\bibinfo {volume} {75}},\ \bibinfo {pages}
  {104305} (\bibinfo {year} {2007})}\BibitemShut {NoStop}%
\bibitem [{\citenamefont {Hong}\ \emph {et~al.}(2022)\citenamefont {Hong},
  \citenamefont {Xiao}, \citenamefont {Hu}, \citenamefont {Ji},\ and\
  \citenamefont {Ran}}]{FTN}%
  \BibitemOpen
  \bibfield  {author} {\bibinfo {author} {\bibfnamefont {Rui}\ \bibnamefont
  {Hong}}, \bibinfo {author} {\bibfnamefont {Ya-Xuan}\ \bibnamefont {Xiao}},
  \bibinfo {author} {\bibfnamefont {Jie}\ \bibnamefont {Hu}}, \bibinfo {author}
  {\bibfnamefont {An-Chun}\ \bibnamefont {Ji}}, \ and\ \bibinfo {author}
  {\bibfnamefont {Shi-Ju}\ \bibnamefont {Ran}},\ }\bibfield  {title} {\enquote
  {\bibinfo {title} {Functional tensor network solving many-body
  {Schr\"odinger} equation},}\ }\href {\doibase 10.1103/PhysRevB.105.165116}
  {\bibfield  {journal} {\bibinfo  {journal} {Physical Review B}\ }\textbf
  {\bibinfo {volume} {105}},\ \bibinfo {pages} {165116} (\bibinfo {year}
  {2022})}\BibitemShut {NoStop}%
\bibitem [{\citenamefont {Mueller}\ \emph {et~al.}(2024)\citenamefont
  {Mueller}, \citenamefont {Zhao}, \citenamefont {Badia},\ and\ \citenamefont
  {Cui}}]{MathMPS1}%
  \BibitemOpen
  \bibfield  {author} {\bibinfo {author} {\bibfnamefont {Nicholas}\
  \bibnamefont {Mueller}}, \bibinfo {author} {\bibfnamefont {Yiran}\
  \bibnamefont {Zhao}}, \bibinfo {author} {\bibfnamefont {Santiago}\
  \bibnamefont {Badia}}, \ and\ \bibinfo {author} {\bibfnamefont {Tiangang}\
  \bibnamefont {Cui}},\ }\bibfield  {title} {\enquote {\bibinfo {title} {A
  tensor-train reduced basis solver for parameterized partial differential
  equations},}\ }\href {\doibase 10.48550/arXiv.2412.14460} {\  (\bibinfo
  {year} {2024}),\ 10.48550/arXiv.2412.14460}\BibitemShut {NoStop}%
\bibitem [{\citenamefont {Chen}\ and\ \citenamefont {Khoo}(2023)}]{MathMPS2}%
  \BibitemOpen
  \bibfield  {author} {\bibinfo {author} {\bibfnamefont {Yian}\ \bibnamefont
  {Chen}}\ and\ \bibinfo {author} {\bibfnamefont {Yuehaw}\ \bibnamefont
  {Khoo}},\ }\href {\doibase 10.48550/arXiv.2305.17884} {\enquote {\bibinfo
  {title} {Combining {Monte} {Carlo} and {Tensor}-network {Methods} for
  {Partial} {Differential} {Equations} via {Sketching}},}\ } (\bibinfo {year}
  {2023})\BibitemShut {NoStop}%
\bibitem [{\citenamefont {Chiara}\ and\ \citenamefont
  {Sanpera}(2018)}]{MathMPS3}%
  \BibitemOpen
  \bibfield  {author} {\bibinfo {author} {\bibfnamefont {Gabriele~De}\
  \bibnamefont {Chiara}}\ and\ \bibinfo {author} {\bibfnamefont {Anna}\
  \bibnamefont {Sanpera}},\ }\bibfield  {title} {\enquote {\bibinfo {title}
  {Genuine quantum correlations in quantum many-body systems: a review of
  recent progress},}\ }\href {\doibase 10.1088/1361-6633/aabf61} {\bibfield
  {journal} {\bibinfo  {journal} {Reports on Progress in Physics}\ }\textbf
  {\bibinfo {volume} {81}},\ \bibinfo {pages} {074002} (\bibinfo {year}
  {2018})}\BibitemShut {NoStop}%
\bibitem [{\citenamefont {Beygi}\ \emph {et~al.}(2015)\citenamefont {Beygi},
  \citenamefont {Klevansky},\ and\ \citenamefont {Bender}}]{coupled}%
  \BibitemOpen
  \bibfield  {author} {\bibinfo {author} {\bibfnamefont {Alireza}\ \bibnamefont
  {Beygi}}, \bibinfo {author} {\bibfnamefont {S.~P.}\ \bibnamefont
  {Klevansky}}, \ and\ \bibinfo {author} {\bibfnamefont {Carl~M.}\ \bibnamefont
  {Bender}},\ }\bibfield  {title} {\enquote {\bibinfo {title} {Coupled
  {Oscillator} {Systems} {Having} {Partial} {PT} {Symmetry}},}\ }\href
  {\doibase 10.1103/PhysRevA.91.062101} {\bibfield  {journal} {\bibinfo
  {journal} {Physical Review A}\ }\textbf {\bibinfo {volume} {91}},\ \bibinfo
  {pages} {062101} (\bibinfo {year} {2015})}\BibitemShut {NoStop}%
\bibitem [{\citenamefont {Andersson}\ \emph {et~al.}(1999)\citenamefont
  {Andersson}, \citenamefont {Boman},\ and\ \citenamefont {\"Ostlund}}]{corr}%
  \BibitemOpen
  \bibfield  {author} {\bibinfo {author} {\bibfnamefont {Martin}\ \bibnamefont
  {Andersson}}, \bibinfo {author} {\bibfnamefont {Magnus}\ \bibnamefont
  {Boman}}, \ and\ \bibinfo {author} {\bibfnamefont {Stellan}\ \bibnamefont
  {\"Ostlund}},\ }\bibfield  {title} {\enquote {\bibinfo {title}
  {Density-matrix renormalization group for a gapless system of free
  fermions},}\ }\href {\doibase 10.1103/PhysRevB.59.10493} {\bibfield
  {journal} {\bibinfo  {journal} {Physical Review B}\ }\textbf {\bibinfo
  {volume} {59}},\ \bibinfo {pages} {10493--10503} (\bibinfo {year}
  {1999})}\BibitemShut {NoStop}%
\bibitem [{\citenamefont {Calabrese}\ and\ \citenamefont
  {Cardy}(2004)}]{field_theory}%
  \BibitemOpen
  \bibfield  {author} {\bibinfo {author} {\bibfnamefont {Pasquale}\
  \bibnamefont {Calabrese}}\ and\ \bibinfo {author} {\bibfnamefont {John}\
  \bibnamefont {Cardy}},\ }\bibfield  {title} {\enquote {\bibinfo {title}
  {Entanglement entropy and quantum field theory},}\ }\href {\doibase
  10.1088/1742-5468/2004/06/P06002} {\bibfield  {journal} {\bibinfo  {journal}
  {Journal of Statistical Mechanics: Theory and Experiment}\ }\textbf {\bibinfo
  {volume} {2004}},\ \bibinfo {pages} {P06002} (\bibinfo {year}
  {2004})}\BibitemShut {NoStop}%
\bibitem [{\citenamefont {Nezhadhaghighi}\ and\ \citenamefont
  {Rajabpour}(2013)}]{harmonicCFT}%
  \BibitemOpen
  \bibfield  {author} {\bibinfo {author} {\bibfnamefont {M.~Ghasemi}\
  \bibnamefont {Nezhadhaghighi}}\ and\ \bibinfo {author} {\bibfnamefont
  {M.~A.}\ \bibnamefont {Rajabpour}},\ }\bibfield  {title} {\enquote {\bibinfo
  {title} {Quantum entanglement entropy and classical mutual information in
  long-range harmonic oscillators},}\ }\href {\doibase
  10.1103/PhysRevB.88.045426} {\bibfield  {journal} {\bibinfo  {journal}
  {Physical Review B}\ }\textbf {\bibinfo {volume} {88}},\ \bibinfo {pages}
  {045426} (\bibinfo {year} {2013})}\BibitemShut {NoStop}%
\bibitem [{\citenamefont {P.~Di~Francesco}\ and\ \citenamefont
  {Sénéchal}(1999)}]{CFT1}%
  \BibitemOpen
  \bibfield  {author} {\bibinfo {author} {\bibfnamefont {P.~Mathieu}\
  \bibnamefont {P.~Di~Francesco}}\ and\ \bibinfo {author} {\bibfnamefont
  {D.}~\bibnamefont {Sénéchal}},\ }\href@noop {} {\emph {\bibinfo {title}
  {Conformal Field Theory}}}\ (\bibinfo  {publisher} {Springer, Heidelberg},\
  \bibinfo {year} {1999})\BibitemShut {NoStop}%
\bibitem [{SM()}]{SM}%
  \BibitemOpen
  \href@noop {} {}\bibinfo {note} {See the Supplemental Material at [with the
  url provided by the publisher] for more details on the infinite functional
  MPS representation of continuous-space many-body wave-functions and the iTEBD
  algorithm based on such a representation.}\BibitemShut {Stop}%
\bibitem [{\citenamefont {Cirac}\ \emph {et~al.}(2021)\citenamefont {Cirac},
  \citenamefont {P\'erez-Garc\'{\i}a}, \citenamefont {Schuch},\ and\
  \citenamefont {Verstraete}}]{mpsreview}%
  \BibitemOpen
  \bibfield  {author} {\bibinfo {author} {\bibfnamefont {J.~Ignacio}\
  \bibnamefont {Cirac}}, \bibinfo {author} {\bibfnamefont {David}\ \bibnamefont
  {P\'erez-Garc\'{\i}a}}, \bibinfo {author} {\bibfnamefont {Norbert}\
  \bibnamefont {Schuch}}, \ and\ \bibinfo {author} {\bibfnamefont {Frank}\
  \bibnamefont {Verstraete}},\ }\bibfield  {title} {\enquote {\bibinfo {title}
  {Matrix product states and projected entangled pair states: Concepts,
  symmetries, theorems},}\ }\href {\doibase 10.1103/RevModPhys.93.045003}
  {\bibfield  {journal} {\bibinfo  {journal} {Reviews of Modern Physics}\
  }\textbf {\bibinfo {volume} {93}},\ \bibinfo {pages} {045003} (\bibinfo
  {year} {2021})}\BibitemShut {NoStop}%
\end{thebibliography}
